\documentstyle[12pt]{article}
\newcommand{\nc}{\newcommand}
\nc{\renc}{\renewcommand}

\nc{\half}{{\textstyle{1\over2}}}
\nc{\etal}{\mbox{\it et al. }}
\nc{\ie}{{\it i.e.}}
\nc{\eg}{{\it e.g.}}

\renc{\thefootnote}{\arabic{footnote}}
\nc{\capt}[1]{{\bf Figure.} {\small\sl #1}}

\nc{\eqs}[2]{\mbox{Eqs.~(\ref{#1},\,\ref{#2})}}
\nc{\eq}[1]{\mbox{Eq.~(\ref{#1})}}

\nc{\figs}[2]{\mbox{Figs.~(\ref{#1},\,\ref{#2})}}
\nc{\fig}[1]{\mbox{Fig~.(\ref{#1})}}

\nc{\tag}[1]{\label{#1} \marginpar{{\footnotesize #1}}}
\nc{\mtag}[1]{\label{#1} \mbox{\marginpar{{\footnotesize #1}}}}
\renc{\baselinestretch}{1.2}
\newlength{\overeqskip}
\newlength{\undereqskip}
\nc{\be}[1]{\begin{equation} \mbox{$\label{#1}$}}
\nc{\bea}[1]{\begin{eqnarray} \mbox{$\label{#1}$}}
\nc{\Section}[2]{\section{#2}\label{#1}}
\nc{\Bibitem}[1]{\bibitem{#1}}
\nc{\Label}[1]{\label{#1}}

\nc{\eea}{\vspace{\undereqskip}\end{eqnarray}}
\nc{\ee}{\vspace{\undereqskip}\end{equation}}
\nc{\bdm}{\begin{displaymath}}
\nc{\edm}{\end{displaymath}}
\nc{\dpsty}{\displaystyle}
\nc{\bc}{\begin{center}}
\nc{\ec}{\end{center}}
\nc{\ba}{\begin{array}}
\nc{\ea}{\end{array}}
\nc{\bab}{\begin{abstract}}
\nc{\eab}{\end{abstract}}
\nc{\btab}{\begin{tabular}}
\nc{\etab}{\end{tabular}}
\nc{\bit}{\begin{itemize}}
\nc{\eit}{\end{itemize}}
\nc{\ben}{\begin{enumerate}}
\nc{\een}{\end{enumerate}}
\nc{\bfig}{\begin{figure}}
\nc{\efig}{\end{figure}}
\nc{\arreq}{&\!=\!&}
\nc{\arrmi}{&\!-\!&}
\nc{\arrpl}{&\!+\!&}
\nc{\arrap}{&\!\!\!\approx\!\!\!&}
\nc{\non}{\nonumber\\*}
\nc{\align}{\!\!\!\!\!\!\!\!&&}

\def\lsim{\; \raise0.3ex\hbox{$<$\kern-0.75em
      \raise-1.1ex\hbox{$\sim$}}\; }
\def\gsim{\; \raise0.3ex\hbox{$>$\kern-0.75em
      \raise-1.1ex\hbox{$\sim$}}\; }
\nc{\DOT}{\hspace{-0.08in}{\bf .}\hspace{0.1in}}
\nc{\Laada}{\hbox {$\sqcap$ \kern -1em $\sqcup$}}
\nc\loota{{\scriptstyle\sqcap\kern-0.55em\hbox{$\scriptstyle\sqcup$}}}
\nc\Loota{{\sqcap\kern-0.65em\hbox{$\sqcup$}}}
\nc\laada{\Loota}
\nc{\qed}{\hskip 3em \hbox{\BOX} \vskip 2ex}

\nc{\real}{{\rm I \! R}}
\nc{\Z}{{\sf Z \!\!\! Z}}
\nc{\complex}{{\rm C\!\!\! {\sf I}\,\,}}
\def\bigid{\leavevmode\hbox{\small1\kern-3.8pt\normalsize1}}
\def\id{\leavevmode\hbox{\small1\kern-3.3pt\normalsize1}}
\nc{\slask}{\!\!\!/}
\nc{\bis}{{\prime\prime}}
\nc{\pa}{\partial}
\nc{\na}{\nabla}
\nc{\ra}{\rangle}
\nc{\la}{\langle}
\nc{\goto}{\rightarrow}
\nc{\swap}{\leftrightarrow}

\nc{\EE}[1]{ \mbox{$\cdot10^{#1}$} }
\nc{\abs}[1]{\left|#1\right|}
\nc{\at}[2]{\left.#1\right|_{#2}}
\nc{\norm}[1]{\|#1\|}
\nc{\abscut}[2]{\Abs{#1}_{\scriptscriptstyle#2}}
\nc{\vek}[1]{{\rm\bf #1}}
\nc{\integral}[2]{\int\limits_{#1}^{#2}}
\nc{\inv}[1]{\frac{1}{#1}}
\nc{\dd}[2]{{{\partial #1}\over{\partial #2}}}
\nc{\ddd}[2]{{{{\partial}^2 #1}\over{\partial {#2}^2}}}
\nc{\dddd}[3]{{{{\partial}^2 #1}\over
        {\partial #2 \partial #3}}}
\nc{\dder}[2]{{{d #1}\over{d #2}}}
\nc{\ddder}[2]{{{d^2 #1}\over{d {#2}^2}}}
\nc{\dddder}[3]{{d^2 #1}\over
        {d #2 d #3}}
\nc{\dx}[1]{d\,^{#1}x}
\nc{\dy}[1]{d\,^{#1}y}
\nc{\dz}[1]{d\,^{#1}z}
\nc{\dl}[1]{\frac{d\,^{#1}l}{(2\pi)^{#1}}}
\nc{\dk}[1]{\frac{d\,^{#1}k}{(2\pi)^{#1}}}
\nc{\dq}[1]{\frac{d\,^{#1}q}{(2\pi)^{#1}}}

\nc{\cc}{\mbox{$c.c.$ }}
\nc{\hc}{\mbox{$h.c.$ }}
\nc{\cf}{cf.\ }
\nc{\erfc}{{\rm erfc}}
\nc{\Tr}{{\rm Tr\,}}
\nc{\tr}{{\rm tr\,}}
\nc{\pol}{{\rm pol}}
\nc{\sign}{{\rm sign}}
\nc{\bfT}{{\bf T }}

\def\GeV{{\rm\ GeV}}

\nc{\cA}{{\cal A}}
\nc{\cB}{{\cal B}}
\nc{\cD}{{\cal D}}
\nc{\cE}{{\cal E}}
\nc{\cG}{{\cal G}}
\nc{\cH}{{\cal H}}
\nc{\cL}{{\cal L}}
\nc{\cO}{{\cal O}}
\nc{\cT}{{\cal T}}
\nc{\cN}{{\cal N}}
\nc{\rvac}[1]{|{\cal O}#1\rangle}
\nc{\lvac}[1]{\langle{\cal O}#1|}
\nc{\rvacb}[1]{|{\cal O}_\beta #1\rangle}
\nc{\lvacb}[1]{\langle{\cal O}_\beta #1 |}
\nc{\bb}{\bar{\beta}}
\nc{\bt}{\tilde{\beta}}
\nc{\ctH}{\tilde{\cal H}}
\nc{\chH}{\hat{\cal H}}
\nc{\1}{\aa}
\nc{\2}{\"{a}}
\nc{\3}{\"{o}}
\nc{\4}{\AA}
\nc{\5}{\"{A}}
\nc{\6}{\"{O}}
\nc{\al}{\alpha}
\nc{\g}{\gamma}
\nc{\Del}{\Delta}
\nc{\e}{\epsilon}
\nc{\eps}{\epsilon}
\nc{\lam}{\lambda}
\nc{\om}{\omega}
\nc{\Om}{\Omega}
\nc{\ve}{\varepsilon}
\nc{\mn}{{\mu\nu}}
\nc{\ka}{\kappa}
\nc{\vp}{\varphi}

\nc{\advp}[3]{{\it  Adv.\ in\ Phys.\ }{{\bf #1} {(#2)} {#3}}}
\nc{\annp}[3]{{\it  Ann.\ Phys.\ (N.Y.)\ }{{\bf #1} {(#2)} {#3}}}
\nc{\apl}[3]{{\it  Appl. Phys. Lett. }{{\bf #1} {(#2)} {#3}}}
\nc{\apj}[3]{{\it  Ap.\ J.\ }{{\bf #1} {(#2)} {#3}}}
\nc{\apjl}[3]{{\it  Ap.\ J.\ Lett.\ }{{\bf #1} {(#2)} {#3}}}
\nc{\app}[3]{{\it Astropart.\ Phys.\ }{{\bf #1} {(#2)} {#3}}}
\nc{\cmp}[3]{{\it  Comm.\ Math.\ Phys.\ }{{ \bf #1} {(#2)} {#3}}}
\nc{\cqg}[3]{{\it  Class.\ Quant.\ Grav.\ }{{\bf #1} {(#2)} {#3}}}
\nc{\epl}[3]{{\it  Europhys.\ Lett.\ }{{\bf #1} {(#2)} {#3}}}
\nc{\ijmp}[3]{{\it Int.\ J.\ Mod.\ Phys.\ }{{\bf #1} {(#2)} {#3}}}
\nc{\ijtp}[3]{{\it Int.\ J.\ Theor.\ Phys.\ }{{\bf #1} {(#2)} {#3}}}
\nc{\jmp}[3]{{\it  J.\ Math.\ Phys.\ }{{ \bf #1} {(#2)} {#3}}}
\nc{\jpa}[3]{{\it  J.\ Phys.\ A\ }{{\bf #1} {(#2)} {#3}}}
\nc{\jpc}[3]{{\it  J.\ Phys.\ C\ }{{\bf #1} {(#2)} {#3}}}
\nc{\jap}[3]{{\it J.\ Appl.\ Phys.\ }{{\bf #1} {(#2)} {#3}}}
\nc{\jpsj}[3]{{\it J.\ Phys.\ Soc.\ Japan\ }{{\bf #1} {(#2)} {#3}}}
\nc{\lmp}[3]{{\it Lett.\ Math.\ Phys.\ }{{\bf #1} {(#2)} {#3}}}
\nc{\mpl}[3]{{\it  Mod.\ Phys.\ Lett.\ }{{\bf #1} {(#2)} {#3}}}
\nc{\ncim}[3]{{\it  Nuov.\ Cim.\ }{{\bf #1} {(#2)} {#3}}}
\nc{\np}[3]{{\it  Nucl.\ Phys.\ }{{\bf #1} {(#2)} {#3}}}
\nc{\pr}[3]{{\it Phys.\ Rev.\ }{{\bf #1} {(#2)} {#3}}}
\nc{\pra}[3]{{\it  Phys.\ Rev.\ A\ }{{\bf #1} {(#2)} {#3}}}
\nc{\prb}[3]{{\it  Phys.\ Rev.\ B\ }{{{\bf #1} {(#2)} {#3}}}}
\nc{\prc}[3]{{\it  Phys.\ Rev.\ C\ }{{\bf #1} {(#2)} {#3}}}
\nc{\prd}[3]{{\it  Phys.\ Rev.\ D\ }{{\bf #1} {(#2)} {#3}}}
\nc{\prl}[3]{{\it Phys.\ Rev.\ Lett.\ }{{\bf #1} {(#2)} {#3}}}
\nc{\pl}[3]{{\it  Phys.\ Lett.\ }{{\bf #1} {(#2)} {#3}}}
\nc{\prep}[3]{{\it Phys\. Rep.\ }{{\bf #1} {(#2)} {#3}}}
\nc{\prsl}[3]{{\it Proc.\ R.\ Soc.\ London\ }{{\bf #1} {(#2)} {#3}}}
\nc{\ptp}[3]{{\it  Prog.\ Theor.\ Phys.\ }{{\bf #1} {(#2)} {#3}}}
\nc{\ptps}[3]{{\it  Prog\ Theor.\ Phys.\ suppl.\ }{{\bf #1} {(#2)} {#3}}}
\nc{\physa}[3]{{\it  Physica\ A\ }{{\bf #1} {(#2)} {#3}}}
\nc{\physb}[3]{{\it  Physica\ B\ }{{\bf #1} {(#2)} {#3}}}
\nc{\phys}[3]{{\it Physica\ }{{\bf #1} {(#2)} {#3}}}
\nc{\rmp}[3]{{\it  Rev.\ Mod.\ Phys.\ }{{\bf #1} {(#2)} {#3}}}
\nc{\rpp}[3]{{\it Rep.\ Prog.\ Phys.\ }{{\bf #1} {(#2)} {#3}}}
\nc{\sjnp}[3]{{\it Sov.\ J.\ Nucl.\ Phys.\ }{{\bf #1} {(#2)} {#3}}}
\nc{\spjetp}[3]{{\it Sov.\ Phys.\ JETP\ }{{\bf #1} {(#2)} {#3}}}
\nc{\yf}[3]{{\it Yad.\ Fiz.\ }{{\bf #1} {(#2)} {#3}}}
\nc{\zetp}[3]{{\it Zh.\ Eksp.\ Teor.\ Fiz.\  }{{\bf #1}  {(#2)} {#3}}}
\nc{\zp}[3]{{\it Z.\ Phys.\ }{{\bf #1} {(#2)} {#3}}}
\nc{\ibid}[3]{{\sl ibid.\ }{{\bf #1} {#2} {#3}}}
\nc{\rf}[1]{(\ref{#1})}
\nc{\nn}{\nonumber \\*}
\nc{\bfB}{\bf{B}}
\nc{\bfv}{\bf{v}}
\nc{\bfx}{\bf{x}}
\nc{\bfy}{\bf{y}}
\nc{\vx}{\vec{x}}
\nc{\vy}{\vec{y}}
\nc{\oB}{\overline{B}}
\nc{\oI}{\overline{I}}
\nc{\oR}{\overline{R}}
\nc{\rar}{\rightarrow}
\nc{\ti}{\times}
\nc{\slsh}{\hskip-5pt/}
\nc{\sm}{Standard~Model~}
\nc{\MP}{M_{\rm Pl}}
\nc{\tp}{t_{\rm Pl}}
\nc{\ave}{\bar{E}}

\nc{\eff}{{\rm eff}}
\nc{\kk}{\vek{k}}
\nc{\pp}{{\rm p}}
\nc{\ga}{g_{a\gamma}}
\nc{\vv}{\\}
\nc{\eee}{{\bf E}}
\nc{\bbb}{{\bf B}}
\nc{\qcd}{T_{\rm QCD}}
\nc{\G}{\rm \ G}
\def\vec#1{{\bf #1}}
\begin{document}

{\title{\vskip-2truecm{\hfill {{\small \\
        }}\vskip 1truecm}
{\bf Magnetic field generation in first order phase transition
bubble collisions}}


{\author{
{\sc Jarkko Ahonen$^{1}$  and Kari Enqvist$^{2}$}\\
{\sl\small Department of Physics, P.O. Box 9,
FIN-00014 University of Helsinki,
Finland}
}
\maketitle
\vspace{2cm}
\begin{abstract}
\noindent
We consider the formation of a ring-like
magnetic field in
collisions of bubbles of broken phase in an abelian
Higgs model.  Particular attention is paid on multiple
collisions. The small collision velocity limit,
appropriate to the electroweak phase transition, is discussed.
We argue that after the completion of the electroweak phase
transition, when averaged over nucleation center distances, 
there exists 
a mean magnetic field $B\simeq 2.0\times 10^{20}$ G with a coherence
length $9.1\times 10^3\GeV^{-1}$ (for $m_H=68$ GeV). 
Because of the ring-like nature of B,
the volume average behaves as $B\sim 1/L$. Taking into account
the turbulent enhancement of the field by inverse cascade,
we estimate that colliding electroweak bubbles would give
rise to a
 mean field $B_{rms}\simeq 10^{-21}$ G at 
10 Mpc comoving scale today.

\end{abstract}
\vfil
\footnoterule
{\small $^1$jtahonen@science.helsinki.fi};
{\small $^2$enqvist@pcu.helsinki.fi};
\thispagestyle{empty}
\newpage
\setcounter{page}{1}

\section{Introduction}
Cosmological first order phase transitions may give rise to
primordial magnetic fields, which  then could act as the  seed field
required for the dynamo explanation of the observed galactic magnetic
fields \cite{dynamo}. A first order phase transition proceeds by nucleation of
bubbles of broken phase in the background of unbroken phase.
In the case of a first order electroweak (EW) phase transition,
the Higgs field inside a given bubble has an arbitrary phase.
The bubbles expand and eventually collide, while new bubbles are
continuously formed, until the phase transition is completed.
This also involves the equilibration of the phases of the complex Higgs
fields, the gradients of which act as a source for gauge fields,
thus making the generations of magnetic fields possible.  
The growth of the bubble is much affected by the non-linear interplay
between the field configuration constituting the bubble and the
background plasma. In the EW case 
hydrodynamical studies show \cite{shocks,hannu} that the expanding bubble is 
preceded by a shock front, which heats up the universe back to the 
critical temperature.  The transition is then completed by 
the merging of the slowly expanding bubbles.

There are two different, but not mutually exclusive, theoretical
scenarios for generating magnetic fields in first order phase transitions.
One employs directly gradients of the complex Higgs field  
in collisions between bubbles of broken phase, as 
discussed in \cite{kv}. The other is based on the spontaneous
appearance of electrical currents and turbulent flow near the
bubble walls and has been applied both to QCD \cite{chengolinto} and EW
\cite{larry,olinto} phase transitions. In the EW phase transition
separation of electric charges occurs  at
the phase boundary because of baryon number gradients. These give rise
to a net current and hence magnetic fields, the fate of which is dependent on
the hydrodynamical details of bubble dynamics. The various dynamical
features have been studied 
carefully in \cite{olinto}, where it was argued that field strengths
of the order of $10^{-29}$ G on a 10 Mpc 
comoving scale could be achieved in EW phase transition
by this mechanism (and in the case of QCD, even larger fields). 
As discussed in \cite{axel}, 
after the phase transition hydromagnetic turbulence
is likely to enhance the seed field by several orders of magnitude, 
thus making the primordial field a plausible candidate for the seed
field.

In the present paper we will focus on magnetic fields created in bubble
collisions, following the treatment of Kibble and Vilenkin \cite{kv},
who showed that in a collision of two bubbles a ring-like magnetic
field is formed. 
(Bubble collisions have also been treated in \cite{hindmarsh},
but mainly with an eye on the defect formation). The starting point is 
an abelian Higgs model, the properties  of which are likely to reflect
the properties of the full EW $SU(2)\times U(1)$ model.
In Section 2 we discuss the collision of the bubbles, and in particular 
the multiple collisions, and show that in all cases the resulting
magnetic field looks qualitatively the same. In Section 3 we introduce
diffusion and consider very slow collision velocities, which are
typical to the EW case. We find out the average field by folding in
the spectrum of separation of nucleation centers and performing the
volume average over the randomly inclined ring-like magnetic field 
configurations.
In Section 4 we present our conclusions.

\section{Bubble collisions}
\subsection{Basic features}
Let us begin by recapitulating some of the features of colliding bubbles
in abelian Higgs model
\cite{kv}. We shall begin by first ignoring diffusion.
The starting point is the U(1)-symmetric lagrangian
\be{lagr}
\cL=-\frac{1}{4}F_{\mu\nu}F^{\mu\nu}+D_{\mu}\Phi (D^{\mu}\Phi )^{\dagger}+
V(\vert\Phi\vert ),
\ee
where $\Phi\equiv\frac{1}{\sqrt{2}}\rho e^{i\Theta}$ 
is the complex Higgs field and the potential
$V$ is assumed to have minima at $\rho=0$ and $\rho=\eta/\sqrt{2}$.
The equations of motion for $\Theta$ and $\rho$ read
\bea{motion2}
\partial_{\mu}\partial^{\mu}\rho -(\partial_{\nu}\Theta -eA_{\nu})^{2}\rho +
2\frac{\partial V}{\partial |\Phi |^{2}}\rho &=&0~ , \cr 
\partial_{\mu}\partial^{\mu}\Theta + e\partial_{\mu}A^{\mu}+2(\partial^{\mu}\Theta
-eA^{\mu})\partial_{\mu}\rho\frac{1}{\rho} &=&0~ . 
\eea
A gauge invariant phase difference can be defined in terms of an
integral over the gradient $D_\mu\Theta$  \cite{kv}.
Before the collision, the phase angle within each bubble may be taken
constant. Following Kibble and Vilenkin \cite{kv} we assume that 
inside the bubble the radial mode
$\rho$ settles rapidly to its equilibrium value $\eta$ and can thus be treated
as a constant. It then follows that a Klein-Gordon equation holds for
$A$ and $\Theta$ separately:
\be{klein}
(\partial_{\mu}\partial^{\mu}+e^2\eta^2 )X=0~,
\ee
where $X=A_\mu\ {\rm or}\ \Theta$. These are not independent but are 
related by virtue of the Maxwell equations:
\bea{maxw}
\partial_{\mu}j^{\mu}  &=&0 , \cr
\partial_{\mu}F^{\mu\nu} &=&j^{\nu} ,
\eea
where $j^{\mu}=-e\rho^2(\partial^{\mu}\Theta + eA^{\mu})$.
The simplest case is that two bubbles nucleate, 
one at $(x,y,z,t)=(0,0,z_0,0)$ and the other at $(x,y,z,t)=(0,0,-z_0,t_{0})$,
and keep expanding with velocity $v$ even after colliding.
Their radii at the collision are denoted by $R_1$ and $R_2$.
It is easy to find out the intersection volume of the bubbles:
\be{intersect}
x^2+y^2=R_{2}^{2}(t)-(z+z_{0})^2 \ ; \ z>0 \ , \\
x^2+y^2=R_{1}^{2}(t)-(z-z_{0})^2 \ ; \ z\leq 0 \ .
\ee
Denoting $R_1(t)=vt$ and $R_{2}(t)=v(t-t_0)$ one may solve for 
the time of intersection:
\be{intersecttime}
t=t_I\equiv
\frac{t_0}{2}+\sqrt{\frac{z_0^2(v^2t_0^2-4z_0^2-4(x^2+y^2))}{v^2(v^2t_0^2-4z_0^2)}}.
\ee
In this Section
 we will always take $v=1$ and $t_{0}=0$ for simplicity,
so that $R_1(t_I)=R_2(t_I)\equiv R$.
As we will discuss later, this is not true for EW phase transition
where  $v\ll 1$,
but nevertheless the assumption  serves a useful illustrative purpose.

Because of the symmetry in the bubble collision, one can
now assume \cite{kv} that the phase angle
$\Theta$ is actually a function of $z$ and $\tau=\sqrt{t^2-x^2-y^2}$.
Also, the symmetry of the problem dictates that in this case the 
electromagnetic potential has the form
$A^{\mu}=x^{\mu}f(\tau ,z)$, 
where $f$ is a function to be determined later. 
One then finds that in the intersection volume the electric and magnetic 
fields read (in cylindrical coordinates)
\bea{kvfields}
\vec{E}&=&-\frac{2tr}{\tau}{\partial\over \partial\tau}
f(\tau ,z)\vec{e}_{r}-
t{\partial\over \partial z}f(\tau ,z)\vec{e}_{z} ~,\cr
\vec{B}&=&r{\partial\over \partial z}f(\tau ,z)\vec{e}_{\Phi}~, 
\eea
where  $r\equiv
\sqrt{x^2+y^2}$.

The solutions for $\Theta$ and $f$ , in the gauge $A^{z}=0$ and
with $\rho =$constant, are obtained  with the initial conditions \cite{kv}
\bea{alku}
\Theta_{|\tau = R}&=&\Theta_{0}\epsilon (z-z_{1})\ ,\ \
\partial_{\tau}\Theta_{|\tau = R}=0,\cr
f_{|\tau= R}&=& 0\ ,\ \
\partial_{\tau}f_{|\tau = R}=\frac{\Theta_0e\eta^2}{R}\epsilon (z-z_1)~,
\eea
and read
\bea{taysi1}
\Theta &=& \frac{\Theta_{0}R}{\pi\tau}\int_{-\infty}^{\infty}\frac{dk}{k}
\sin {k(z-z_1)}
\big[\cos \omega (\tau -R)+\frac{1}{\omega R}\sin \omega (\tau -R)\big]~,\nn
f&=&\frac{\Theta_{0}Re\eta^2}{\pi\tau^3}\int_{-\infty}^{\infty}
\frac{dk}{k}\sin {k(z-z_1)}
\big[\frac{R-\tau}{\omega^2 R}\cos\omega (\tau -R) \cr
&+&\big(\frac{\tau}{\omega}+
\frac{1}{\omega^3 R}\big)
\sin \omega (\tau -R)\big],
\eea
where $\omega\equiv\sqrt{k^2 + e^2\eta^2}$ and $R$ is the 
radius of the bubbles at the collision and $z_1$ the point of first collision
on the z-axis.

In Fig. 1 and 2 we display the time evolution of the
absolute value of the magnetic field and $\Theta$.
One can see that $\vec{B}$ spreads out in the whole intersection
region (which in Fig. 1 and 2 correspond to the range
$50-t\lsim z \lsim 50+t$) and oscillates 
rapidly with increasing frequency as one moves 
from the center of the 
regime to the edge. As time goes on, the amplitude of $\vec{B}$
oscillation decreases as $1/\tau^{2}$ while the frequency increases.
 Therefore the mean 
field at  sufficiently large scales, of the order of $10/(e\eta)$, is 
zero everywhere else except in the 
middle of the collision regime.  
The energy density $\int_{-\infty}^{\infty}B^2dV/V$ can be seen to
scale as $1/t$.

\begin{figure}
\leavevmode
\centering
\vspace*{90mm}
\includegraphics{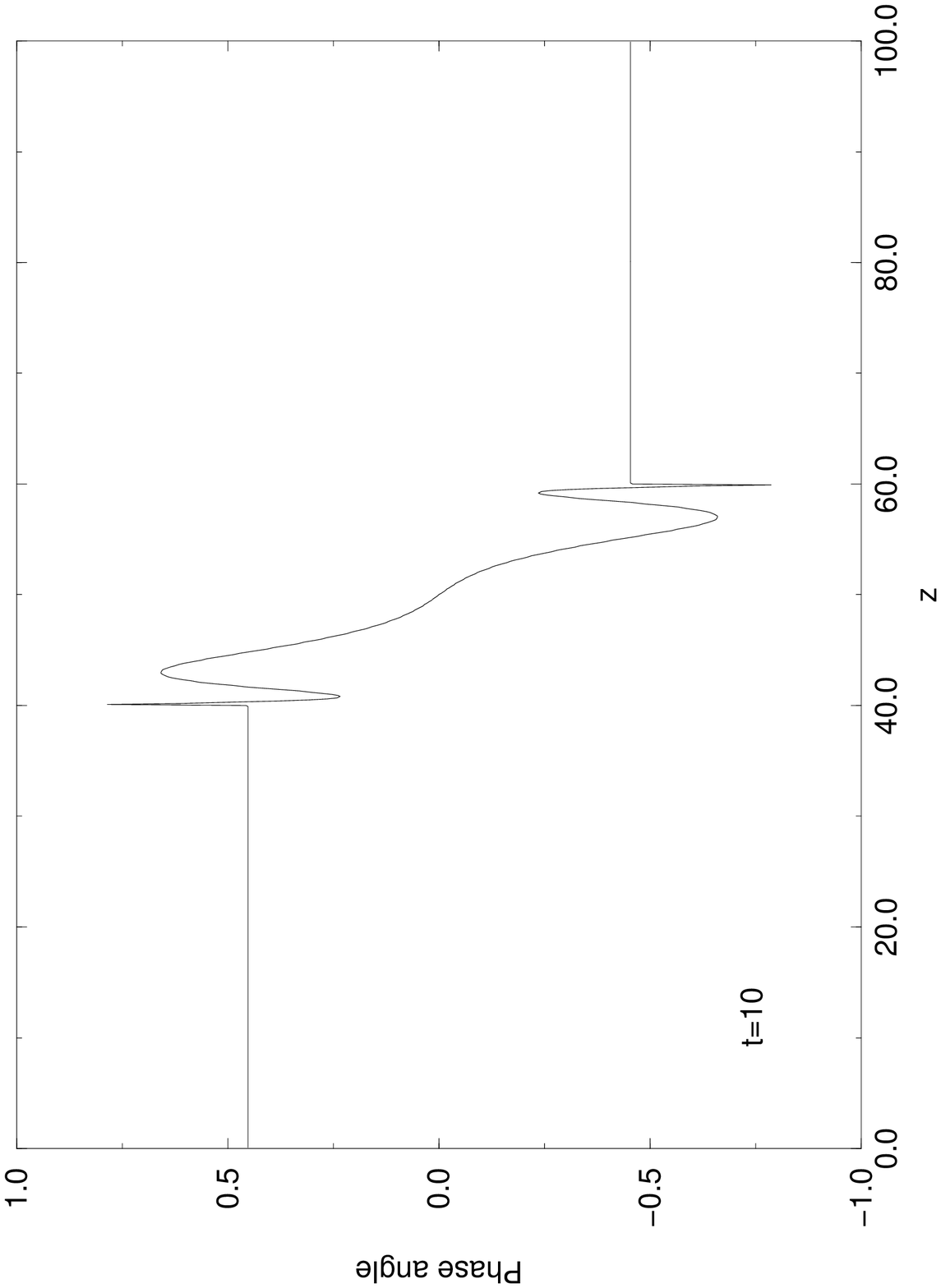}
\includegraphics{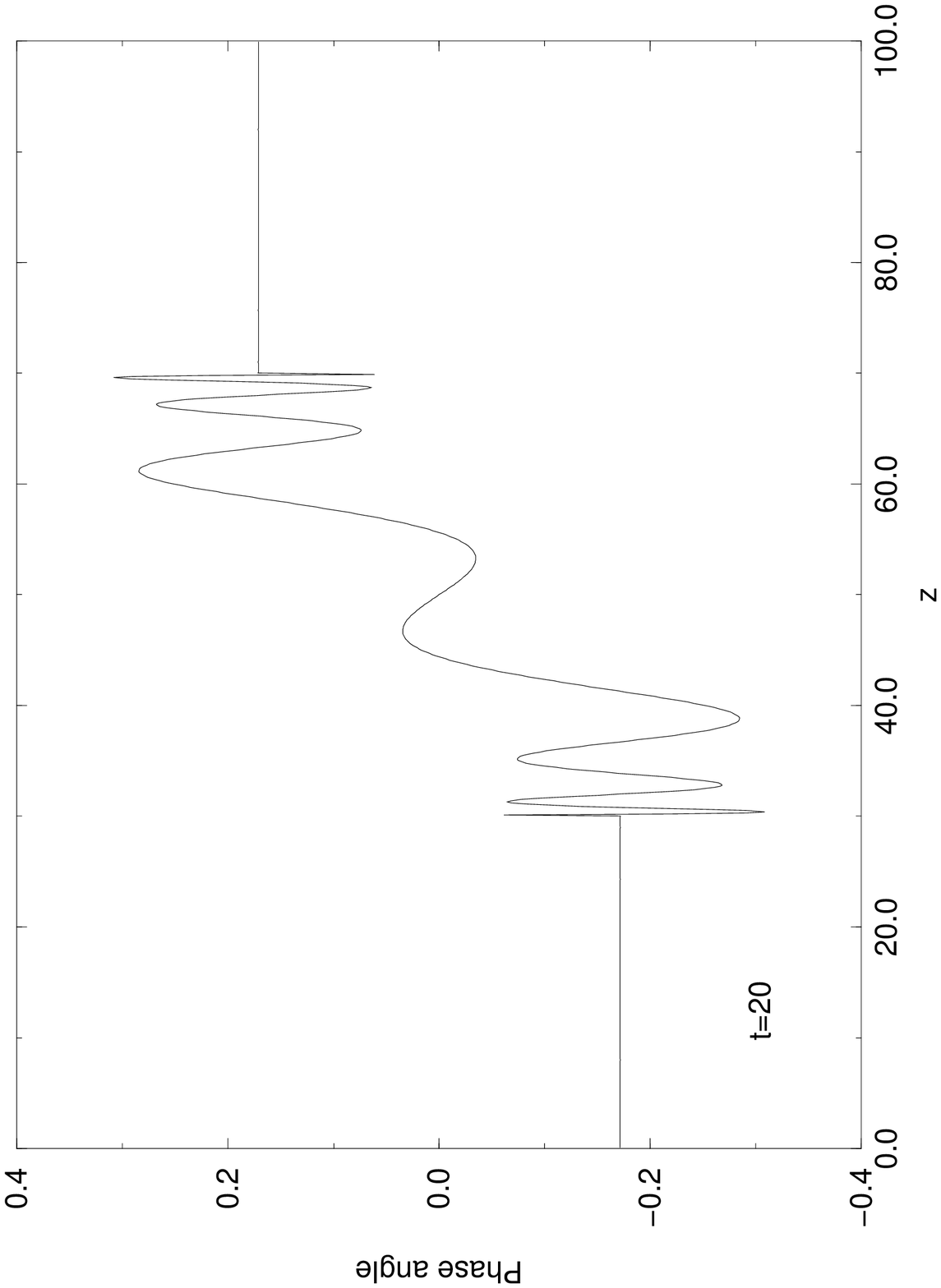}
\includegraphics{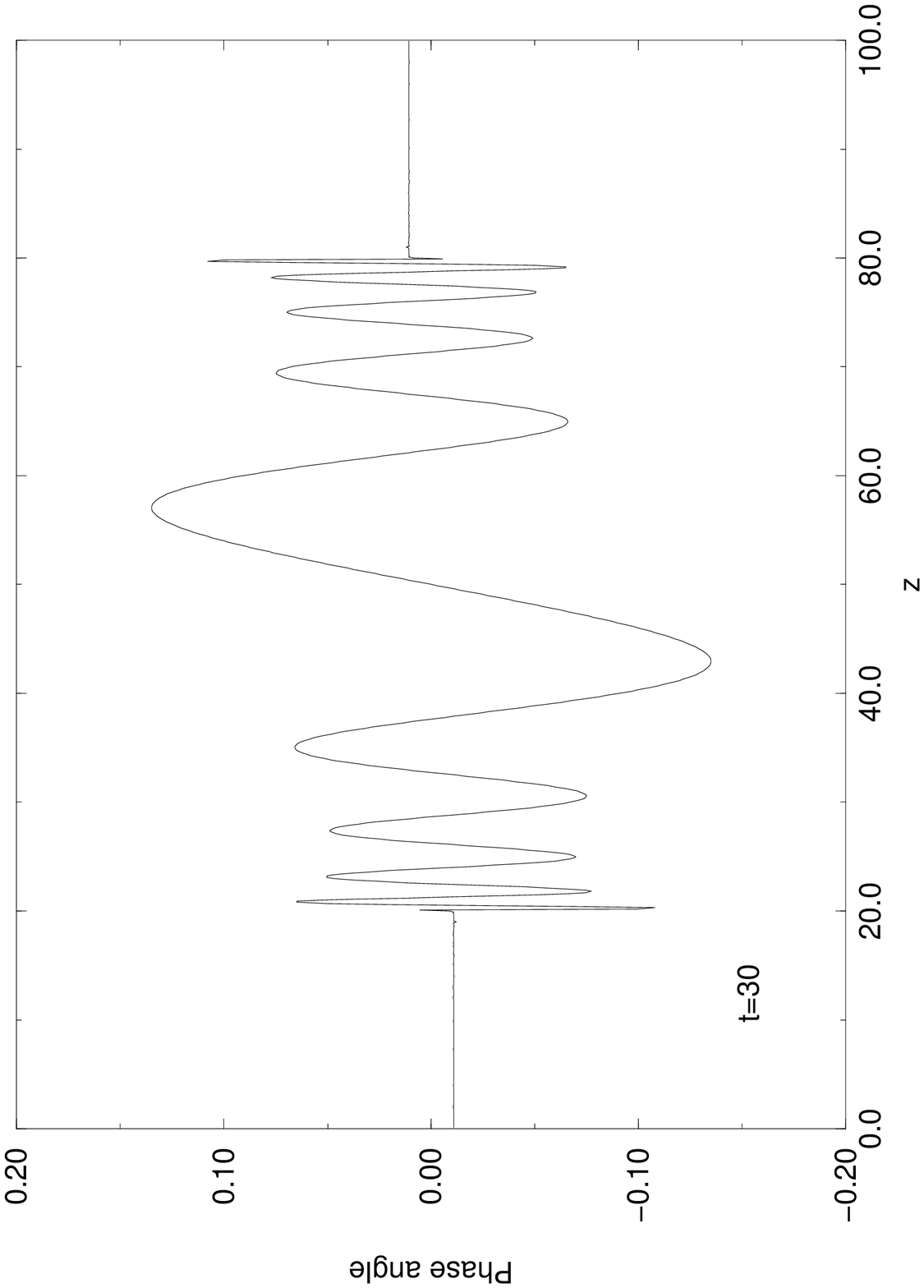}   
\includegraphics{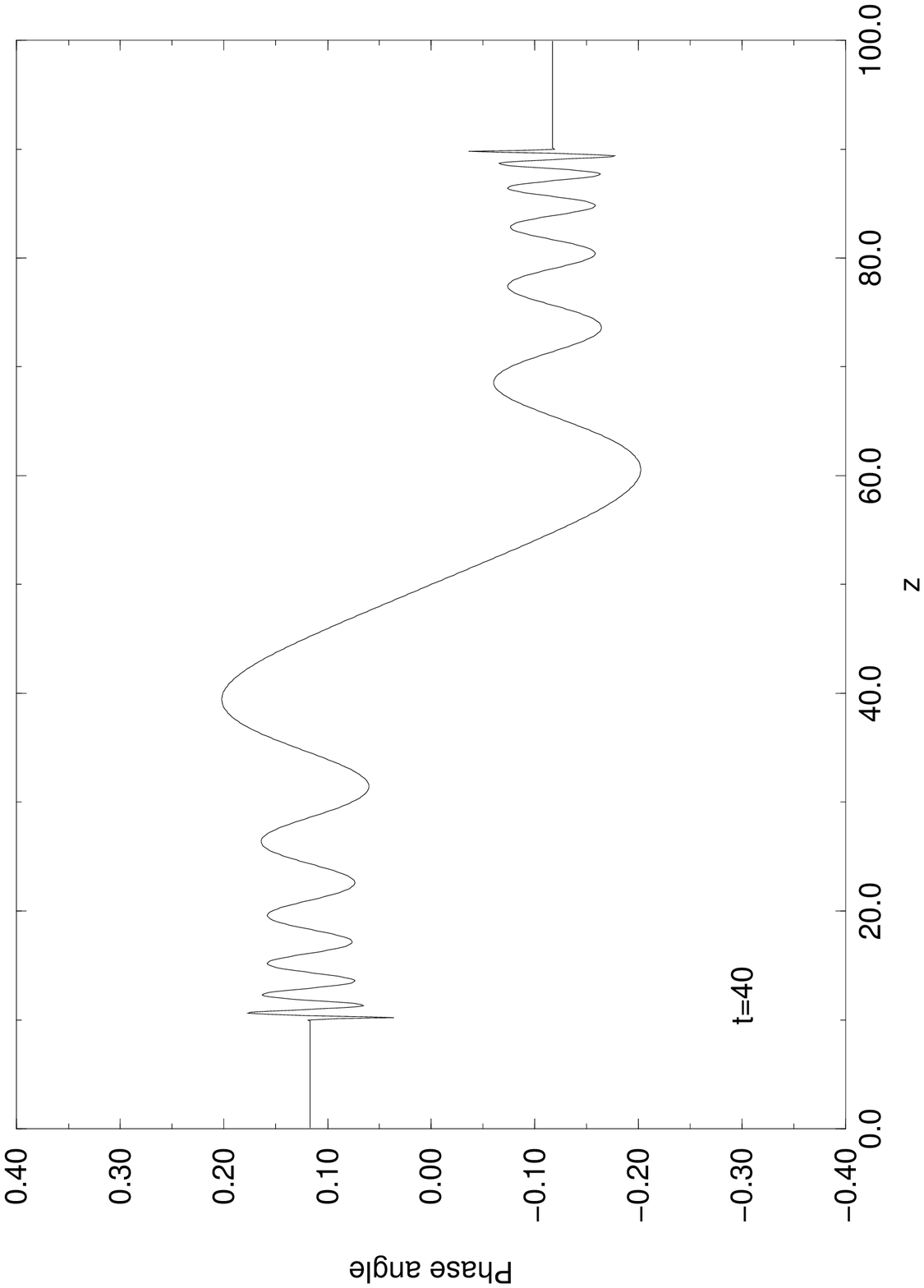}   
\caption{The time evolution of the phase angle
$\Theta$ along the radius $r=1$ for $t$=10, 20, 30 and 40 after the initial
collision. The radius of the bubbles at the collision has been chosen here  
$R=10$, and the collision point on the z-axis is $z_1=50$.  
(The  units are such that $e\eta =1$).}
\label{kuva1}
\end{figure}

\begin{figure}
\leavevmode
\centering
\vspace*{90mm} 
\includegraphics{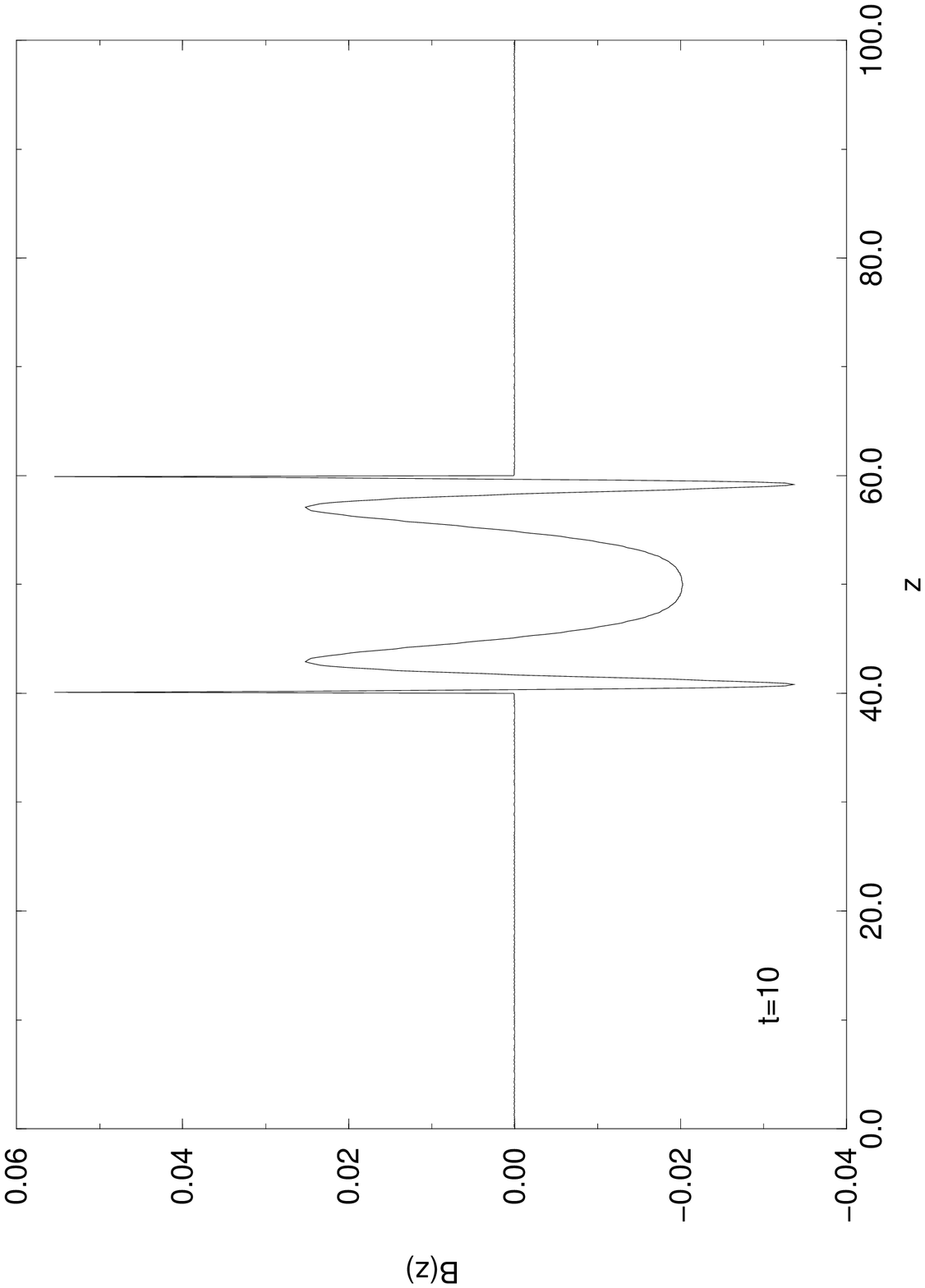}
\includegraphics{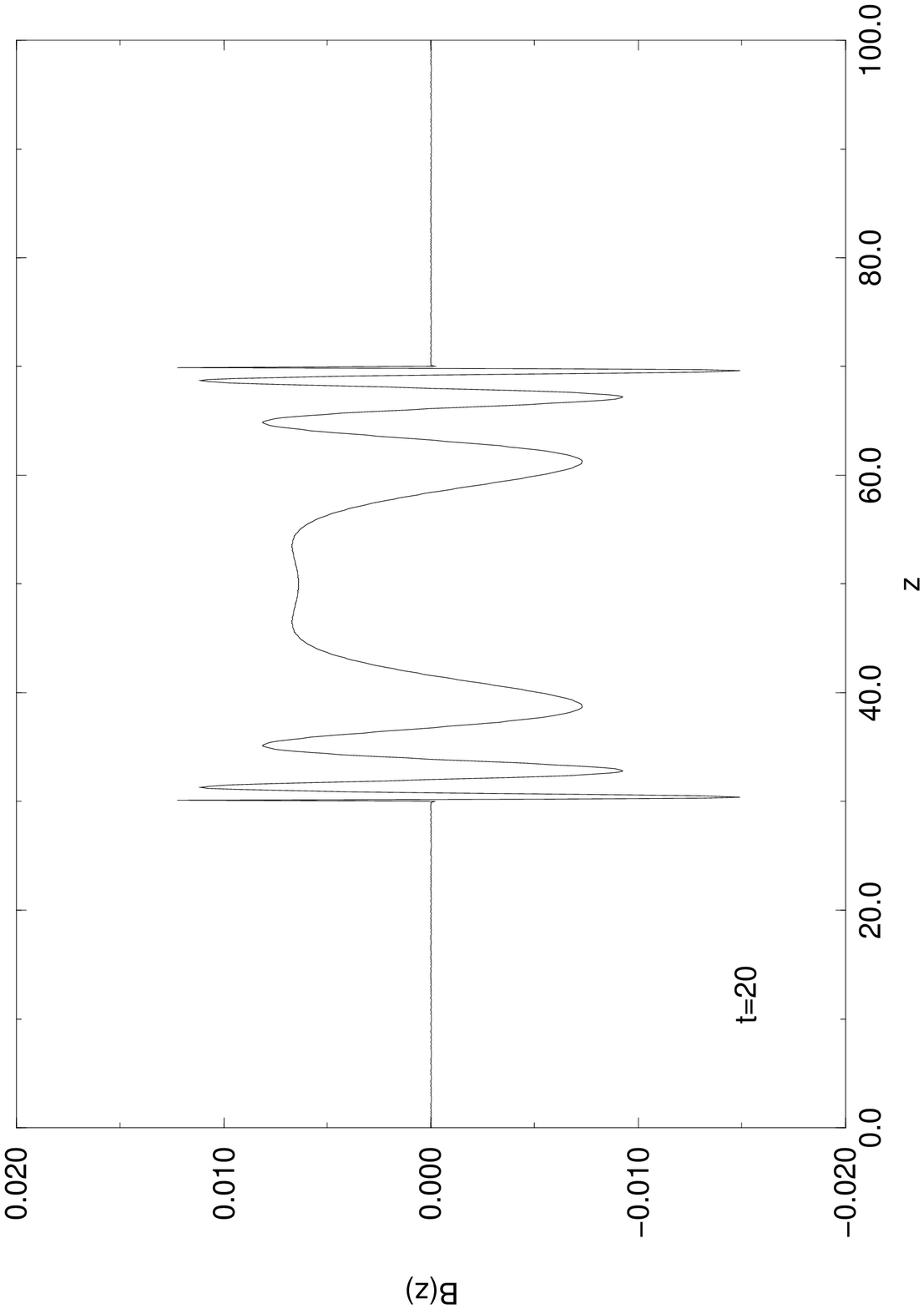}   
\includegraphics{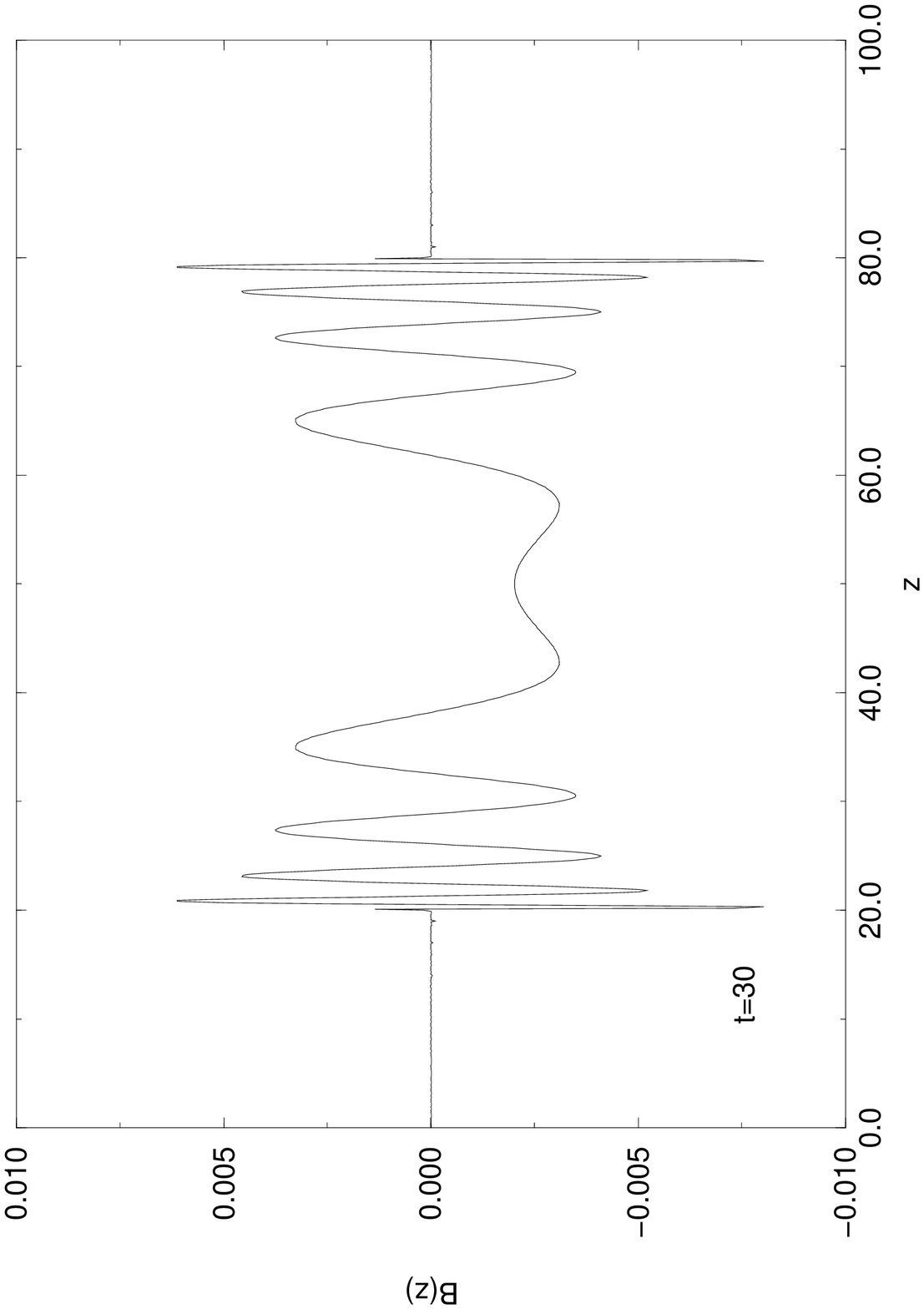}
\includegraphics{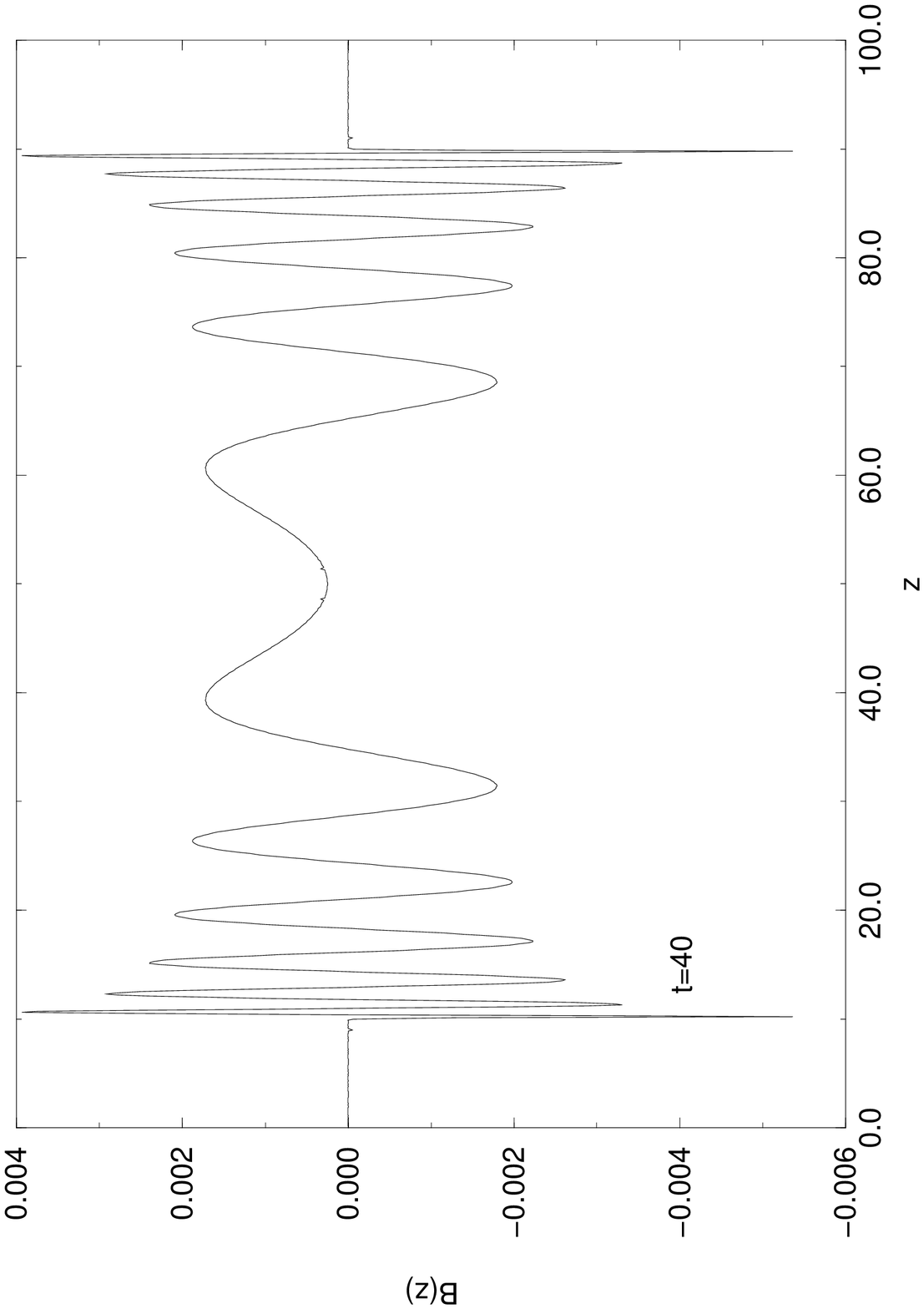}
\caption{The time evolution of 
$B$ along the radius $r=1$ for $t$=10, 20, 30 and 40 after the initial
collision. The radius of the bubbles at the collision has been chosen here 
$R=10$, and the collision point on the z-axis is $z_1=50$.
(The  units are such that $e\eta =1$).}
\label{kuva2}       
\end{figure}

\subsection{Full-empty collisions}
After the bubbles have intersected, they will be subject to new collisions.
Indeed, first order phase transitions are locally completed by merging
of several bubbles.
 Let us assume that a sufficiently long time has elapsed since the first
collision so that we may approximate the result of the collision by a
spherical bubble with radius $R_1$.
It contains a magnetic field and a phase angle
(we shall call it ``full'' bubble), which
are given by
\eq{taysi1}.
Strictly speaking, these solutions cannot be valid inside the whole 
full bubble, but we assume that they are valid in the collision region
where the full bubble merges with another bubble.

When a full bubble collides with a recently nucleated bubble with no
gauge field inside (which we shall call an ``empty'' bubble), some of
symmetries of the empty-empty collision 
remain. Because of the reduced symmetry, we will now take $\tau =\sqrt{
t^2-x^2}$.

The electromagnetic potential in the 
full-empty case can be obtained by approximating the  
initial conditions by
$f_{|\tau =R_1,y=0,z\geq 0}=f_{0}(1-e^{-Cz})(1-e^{-C z^2})^{-1}H(z_2-z)$, 
where $H$ is the 
Heaviside theta-function, $f_0$ the asymptotic form of the electromagnetic
potential and
$\partial_{\tau}f_{|\tau =R_1,y=0,z\geq 0}=0$. $z_2$ is the point of collision
on the z-axis, which is chosen as the symmetry axis of the collision.
These initial conditions should approximate quite well the average behaviour of the gauge
potential in the 'full' bubble, which subsequently collides with an 
empty one assuming that the bubbles are big enough so that the 
vector potential $\vec{A}$ can be taken to point to a constant direction 
in the collision region of the full bubble.

The vector potential is zero at $z=0$ and $A_i=x_if_0$ in 
the asymptotic $z\rightarrow\infty$ region, where $f_0$ is given by
\be{asympta}
f_{0}=f(z\to\infty)
=\frac{\Theta_{0}\eta R}{\tau^3}\Big[\frac{R-\tau}{e\eta R}
\cos (e\eta (\tau -R)) +
\left(\tau + \frac{1}{e\eta R}\right)\sin (e\eta (\tau -R))\Big]~.
\ee
The parameter $C$ can be found by matching the z-derivatives of 
$f_{0}(1-e^{-Cz})(1-e^{-Cz_2})^{-1}$
and \eq{taysi1} at $z=z_1=0$.  

This approximates rather well the overall behaviour of 
$f$, but does not take into account
the rapid oscillations involved. However, when diffusion is included (Sect. 3),
it is precisely these rapid oscillations which get diffused.
An expression for $\vec{A}$, obeying the approximate initial
conditions, can then be found, but 
 it cannot be expressed in a simple form and we do not reproduce it
here. A more tractable 
picture can be obtained by making the assumption that the empty
bubble is much smaller than the full bubble. This also seems 
a natural assumption. We may then write $f=f_0+\delta f$, where 
$\delta f$ is a perturbation and $f_0$ is 
the asymptotic solution obtained from the collision of two empty
bubbles, \eq{asympta}. 

Let us choose a new coordinate system so that
the electromagnetic potential in the full bubble can be written 
in the form
\be{smallb}
\vec{A}=R_{1} f_{0}\sin\alpha\vec{e}_{y}~,
\ee
where $f_{0}$ is the asymptotic form of $f$, \eq{asympta}, $\alpha$ the 
intersection angle and $R_1$ the
radius of the large bubble.
Also, we take $\Theta$ to be at its asymptotic value as
given by 
\be{asymptfe}
\Theta_{0}=\frac{\Theta_{00}R}{\tau}\left(\cos (e\eta (\tau -R))+
\frac{\sin (e\eta (\tau -R))}{e\eta
R}\right)~.
\ee
Here $\Theta_{00}$ is the original phase difference of the two initial 
colliding
empty bubbles, and $R$ refers to the radii of the these bubbles.
The initial conditions can now be written in the form 
\bea{initsmall}
\Theta_{|\tau = R_{1}}&=&\Theta_{0}\epsilon (z-z_{1})+\Theta_{1}\ ,\ \
\partial_{\tau}\Theta_{|\tau = R_{1}}=0,\cr
f_{|\tau= R_{1}}&=& a(1-\epsilon (z-z_1))\ ,\ \
\partial_{\tau}f_{|\tau = R_{1}}=b(1-\epsilon (z-z_1)),
\eea
where $a= {f_0}/{2}$ and $b= -e\eta^2(\Theta_0+{
3f_0}/{2e\eta^2})/{R_1}$.
Here $\tau$ is once again $\tau^2\equiv t^2-x^2-y^2$ and $\omega^2=
k^2+e^2\eta^2$.
One then finds that
\bea{taast}
\Theta &=&\frac{\Theta_{0}R_1}{\pi\tau}\int_{-\infty}^{\infty}
\big[\frac{\pi\Theta_{1}\delta (l)}{
\Theta_{0}}\cos l(z-z_{1})+\frac{\sin l(z-z_{1})}{l}\big]\big[\cos
\big(\sqrt{e^2\eta^2+l^2}(\tau -R_{1})\big)\cr
& &\cr
&+&\frac{1}{\omega R_1}\sin
(\sqrt{e^2\eta^2+l^2}(\tau-R_{1}))\big]dl~,
\eea
and
\bea{taasa}
f&=&\int_{\infty}^{\infty}\frac{-1}{\omega^3\tau^2}\Big( \Big[\cos (
\omega R_1)
\Big( 3a\omega^2R_1+b\omega^2R_{1}^{2}\Big)+\sin (\omega R_1)\Big( 
a\omega^3R_{1}^{2}-3a\omega -b\omega R_1\Big)\Big]\cr
& &\cr
&\times&
\Big[\sin (\omega\tau )+
\frac{\cos (\omega \tau)}{\omega\tau}\Big]+
\frac{R_1}{2(\sin (\omega R_1)-\omega R_1\cos (\omega R_1))}\Big(
(a\omega^4R_{1}^{2}-\cos (2\omega R_1)\cr
& &\cr
&\times&\Big[6a\omega^2+2b\omega^2R_1-
a\omega^4R_{1}^{2}\Big]+\sin (2\omega R_1)\Big[b\omega +\frac{3a\omega}{
R_1}-4a\omega^3R_1-b\omega^3R_{1}^{2}\Big]\Big)\cr
& &\cr
&\times&\Big(-\cos(\omega\tau )+
\frac{\sin (\omega \tau )}{\omega \tau}\Big)\Big)
\Big[\frac{\sin k(z-z_{1})}{\pi k}-
\delta (k)\cos k(z-z_{1})
\Big]dk~.
\eea
From the complicated expression \eq{taasa} one then finds that approximately
\be{approbee}
\vec{B}\simeq R_{1}\sin\alpha (\partial_{z}f)\vec{e}_{x}~.
\ee
To take into account the angle $\alpha$ between $\vec B$ in the 
full bubble and the line of collision, one should substitute
$y\rightarrow y \cos\alpha -z\sin\alpha$ and $z\rightarrow
z\cos\alpha + y\sin\alpha$. Qualitatively, the situation is 
however equivalent to 
the case of empty-empty collision. As can be seen from \eq{approbee},
the resulting magnetic field is again a ring, although it is slanted 
by the angle $\alpha$ with respect to the z-axis.
Its time evolution
is also similar to what is presented in Fig. 1.
Because almost every collision that takes place between empty and full bubbles
can be treated in the approximation above, the 
structure of magnetic fields generated 
will all be essentially equal.

\subsection{Subsequent collisions}
As time goes on, more and more bubbles may have collided at least once,
and the subsequent collisions take place between already ``full'' bubbles,
i.e. bubbles which contain magnetic fields. In such collisions there are
no symmetries left, and the resulting equations of motion are much more
complicated compared with the previous cases. However, in the case where
one of the bubbles is much larger than the other, it is again possible
to approximate the small bubble as a small perturbation on a bubble with
infinite radius. We consider a collision where the magnetic fields
inside the bubbles point to arbitrary directions. The bubbles may again
be taken to collide along the $z$-axis with the gauge potentials
almost constant in the colliding region. 
Because all symmetries are lost, one can not assume
that $A_{i}=x_{i}f$. In general, however, these collisions can be very
well approximated with the initial conditions given by
\eq{initsmall} for $\Theta$ and $\vec{A}$. Therefore,
qualitatively the full-full collision does not differ from the empty-empty
collision discussed in Sect. 2.1. 

Because both $\Theta$ and $\vec A$ obey the Klein-Gordon equation, the
solutions are simple waves. That is why in the case of two colliding bubbles 
the qualitative behaviour of the generated magnetic field is found not 
to depend on whether the bubbles had already collided or not. A more
involved situation arises when one considers a simultaneous collision
of more than two bubbles. The waves can then interfere and produce more
complicated patterns.

To demonstrate this, 
let us assume that 3 (empty) bubbles move along the $z$-axis and
collide simultaneously. We can then use symmetry and again
define $\vec A = \vec x f$. The initial conditions read as
in Sect. 2.1, but now for the three bubbles.
We have solved the equation of motion numerically, and
the results are presented in Fig. 3 and Fig. 4, 
where we show the time evolution
of $\Theta$ and $B$.

\begin{figure}
\leavevmode
\centering
\vspace*{90mm}
\includegraphics{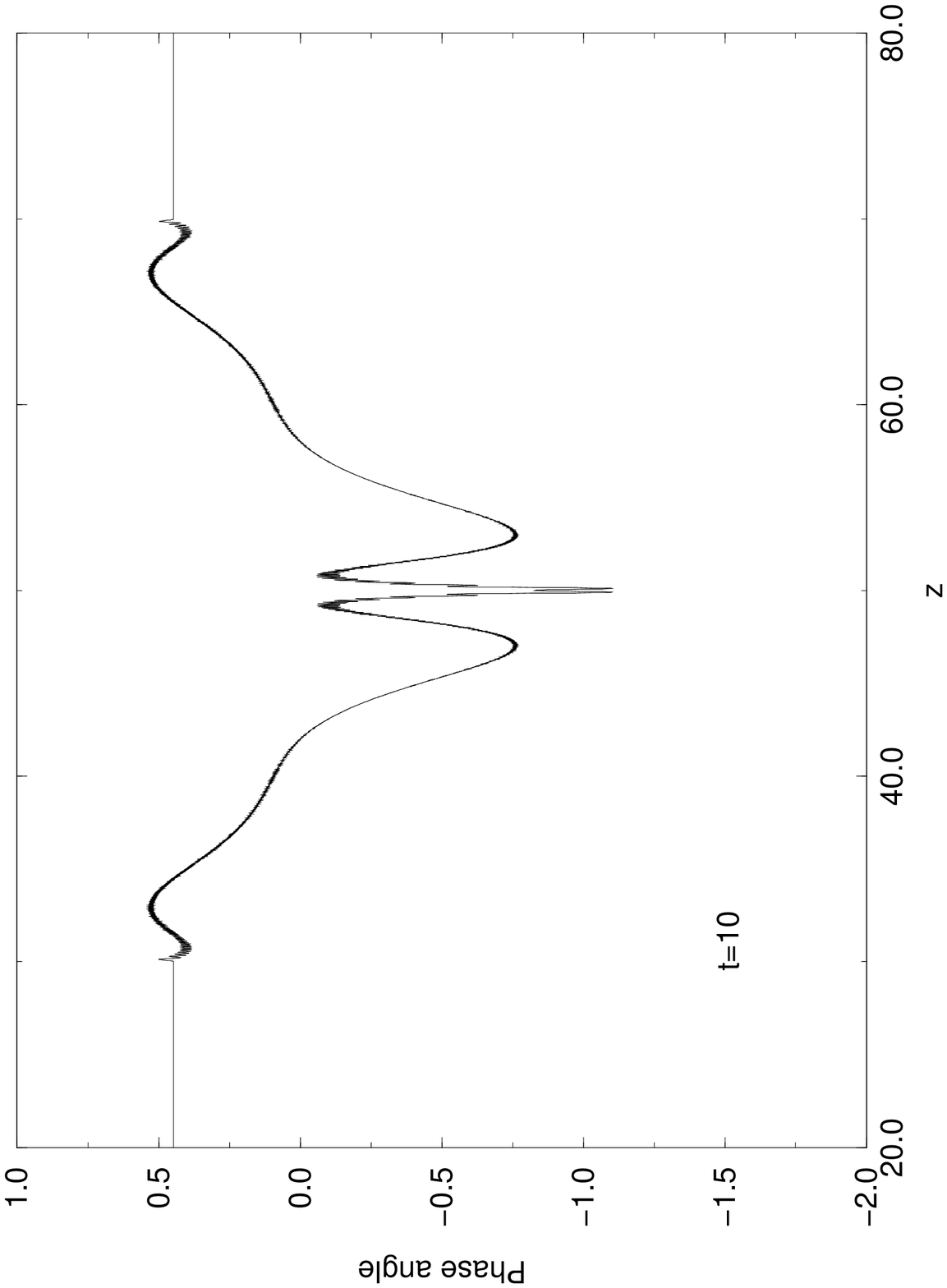}
\includegraphics{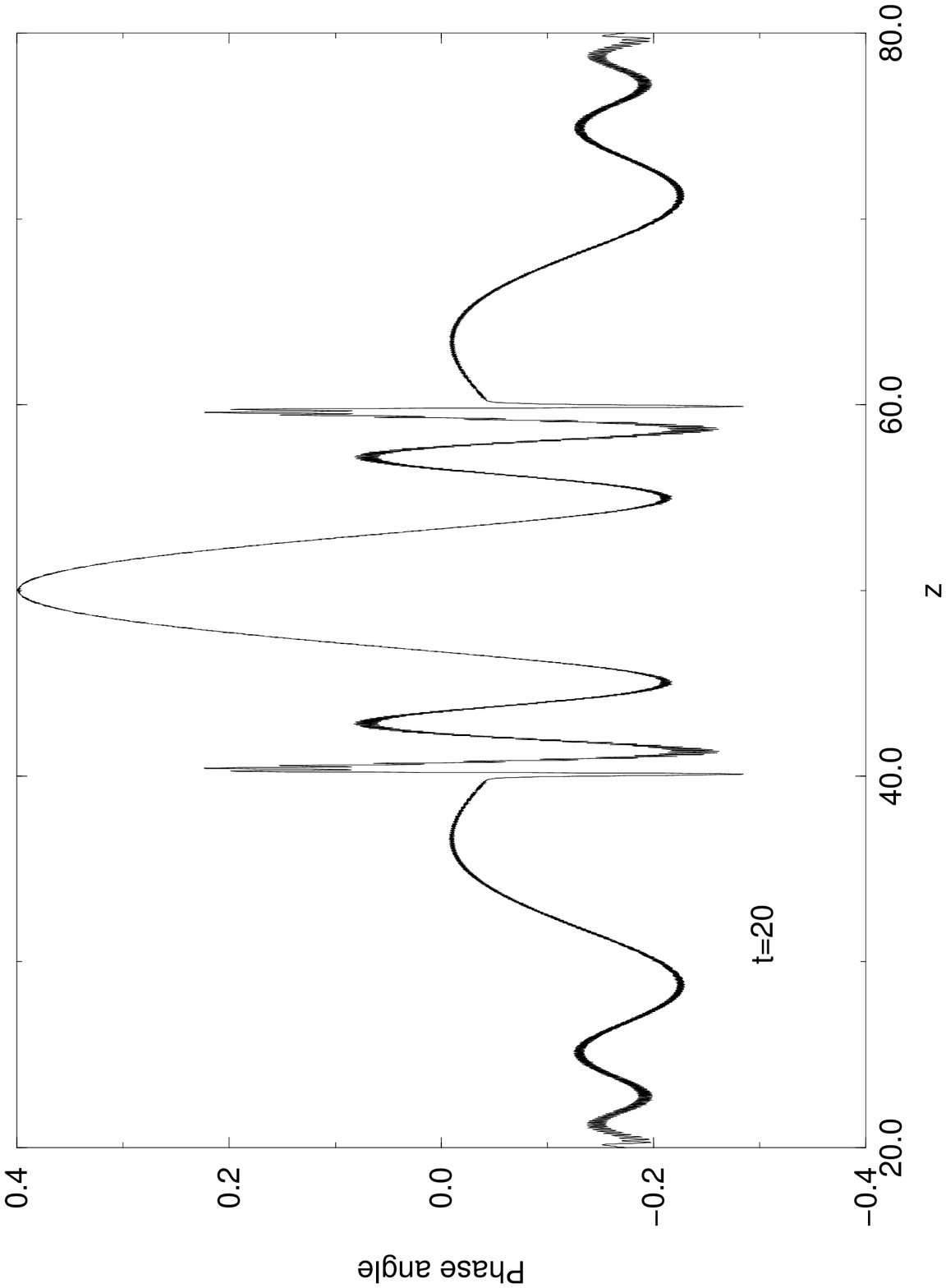}
\includegraphics{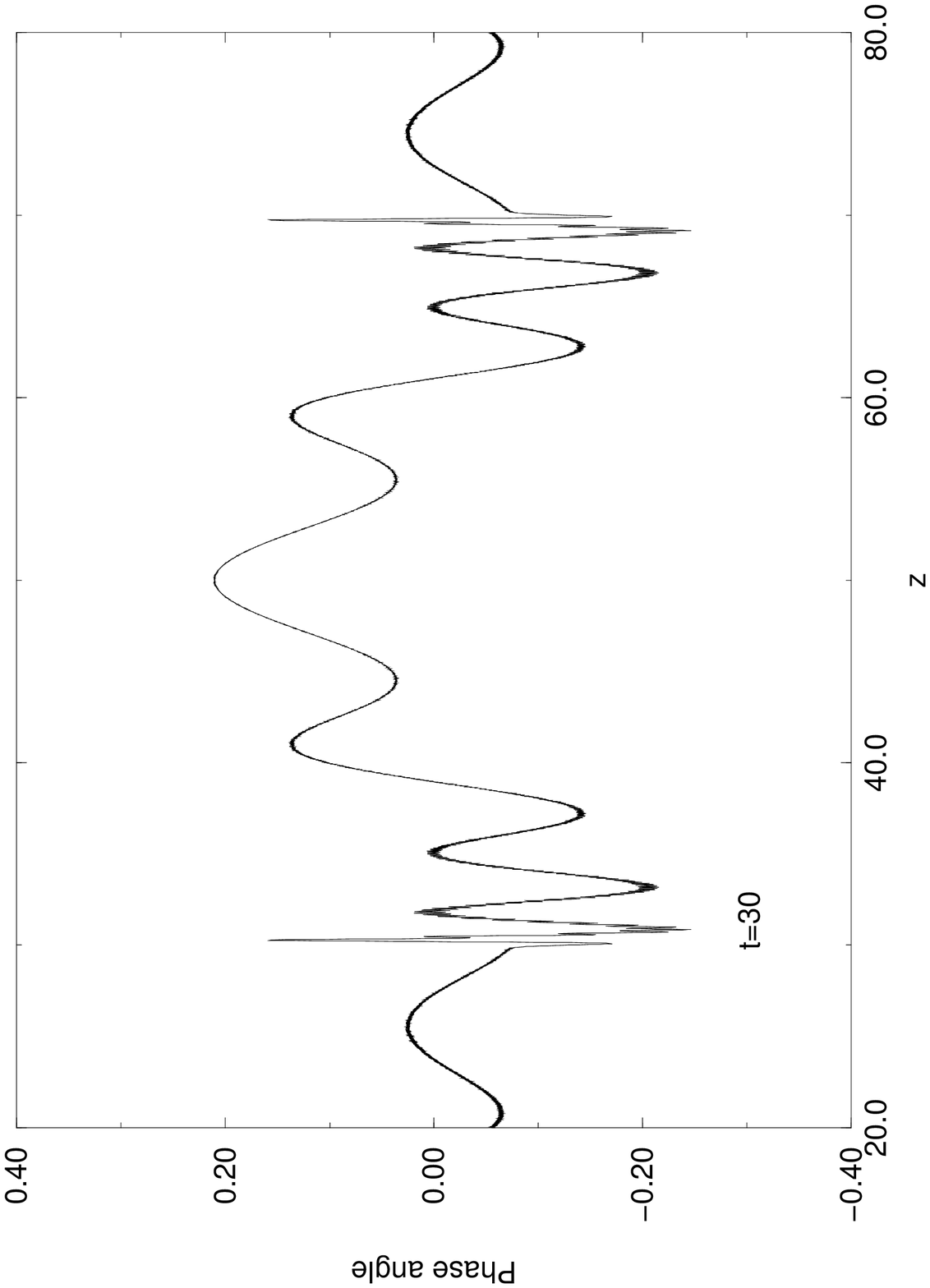}
\includegraphics{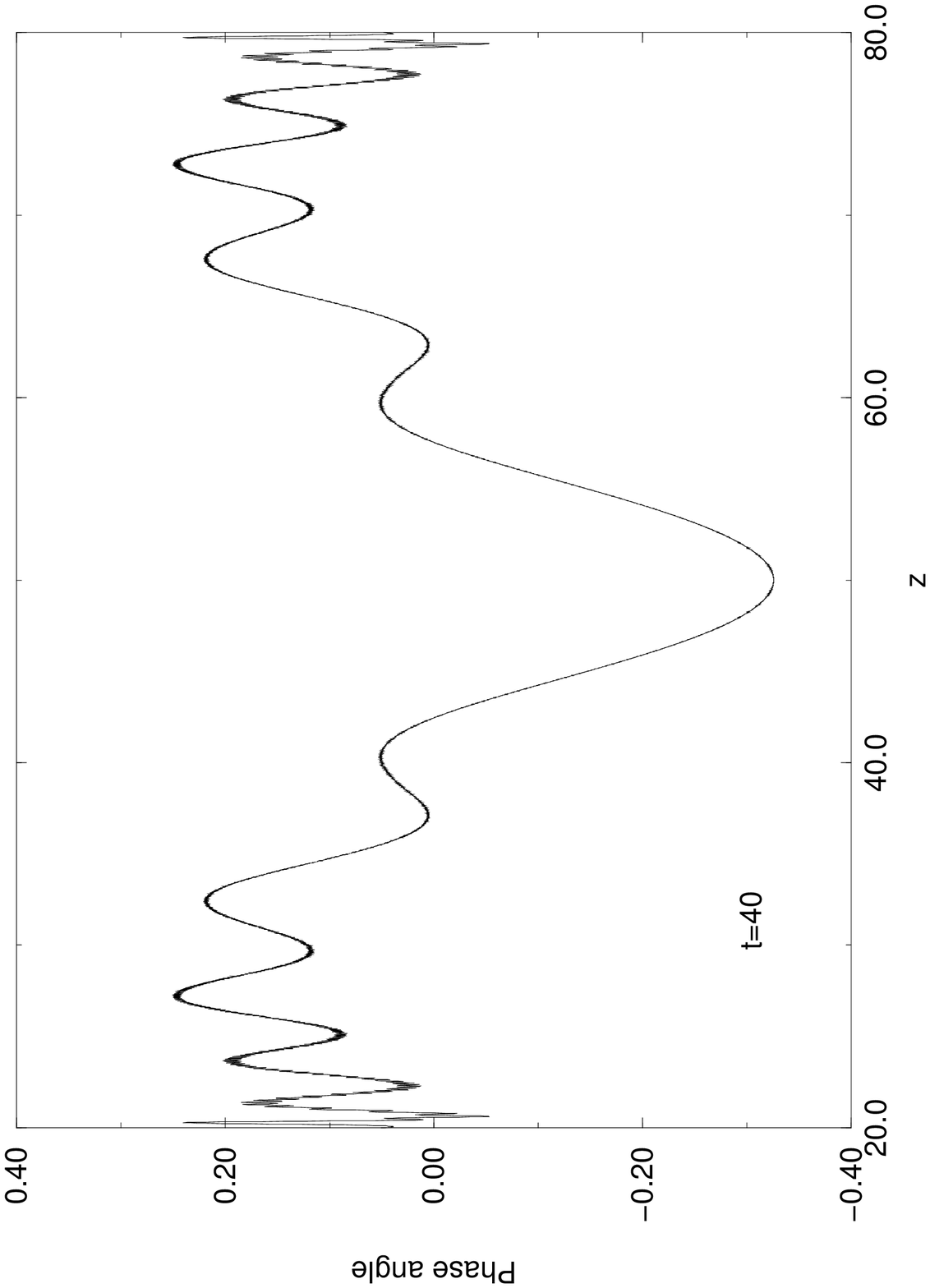}
\caption{The time evolution of
$\Theta$ in symmetric collision of three bubbles each having radius
$R=10$ at the collision. Here  $r=1$ and time refers to the time elapsed
after the initial collision of the bubbles. The points of initial collision
of the bubbles on the z-axis are 40 and 60 (and units are $e\eta=1$).}   
\label{kuva3}
\end{figure}

\begin{figure}
\leavevmode
\centering
\vspace*{90mm}
\includegraphics{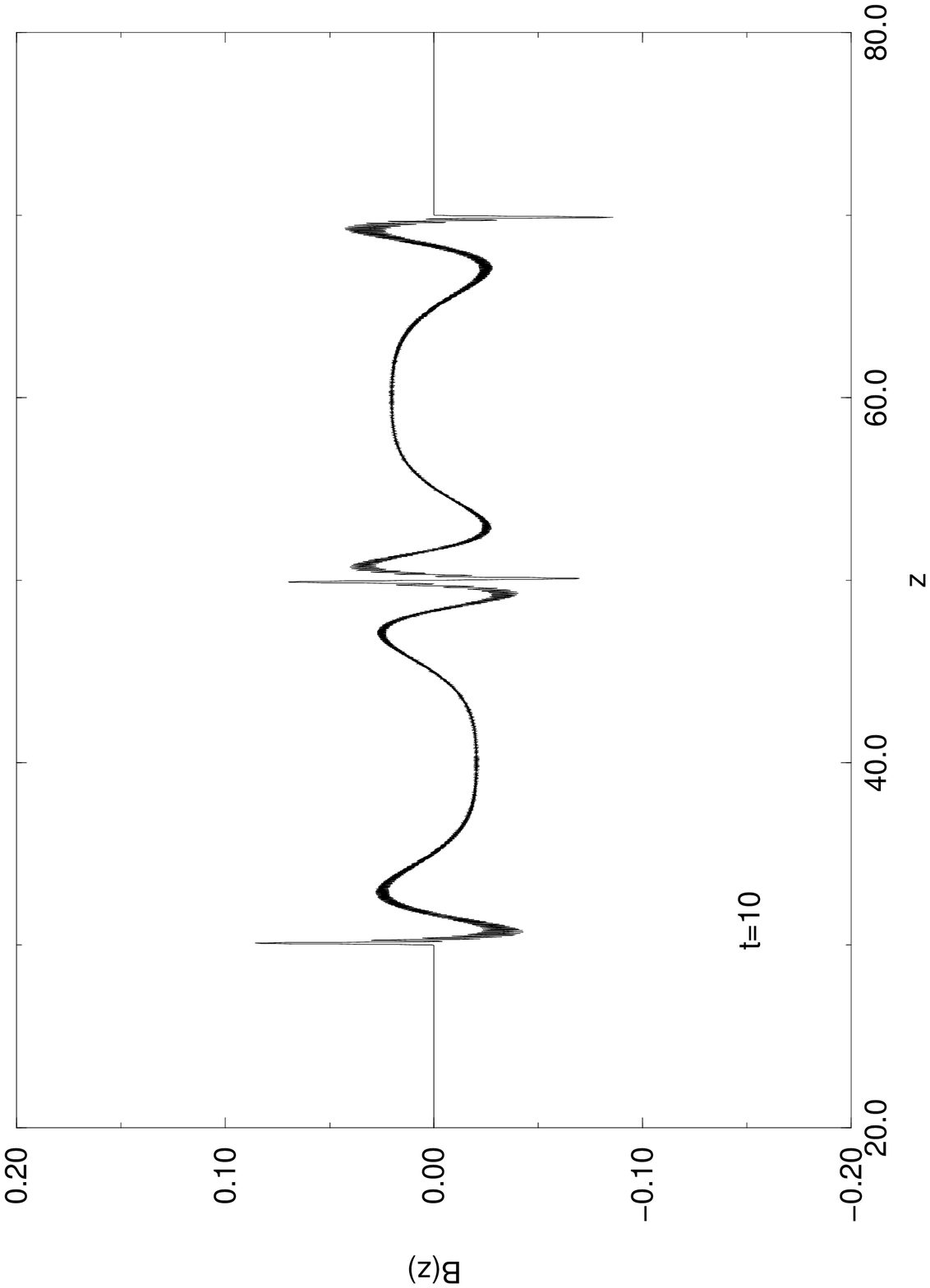}
\includegraphics{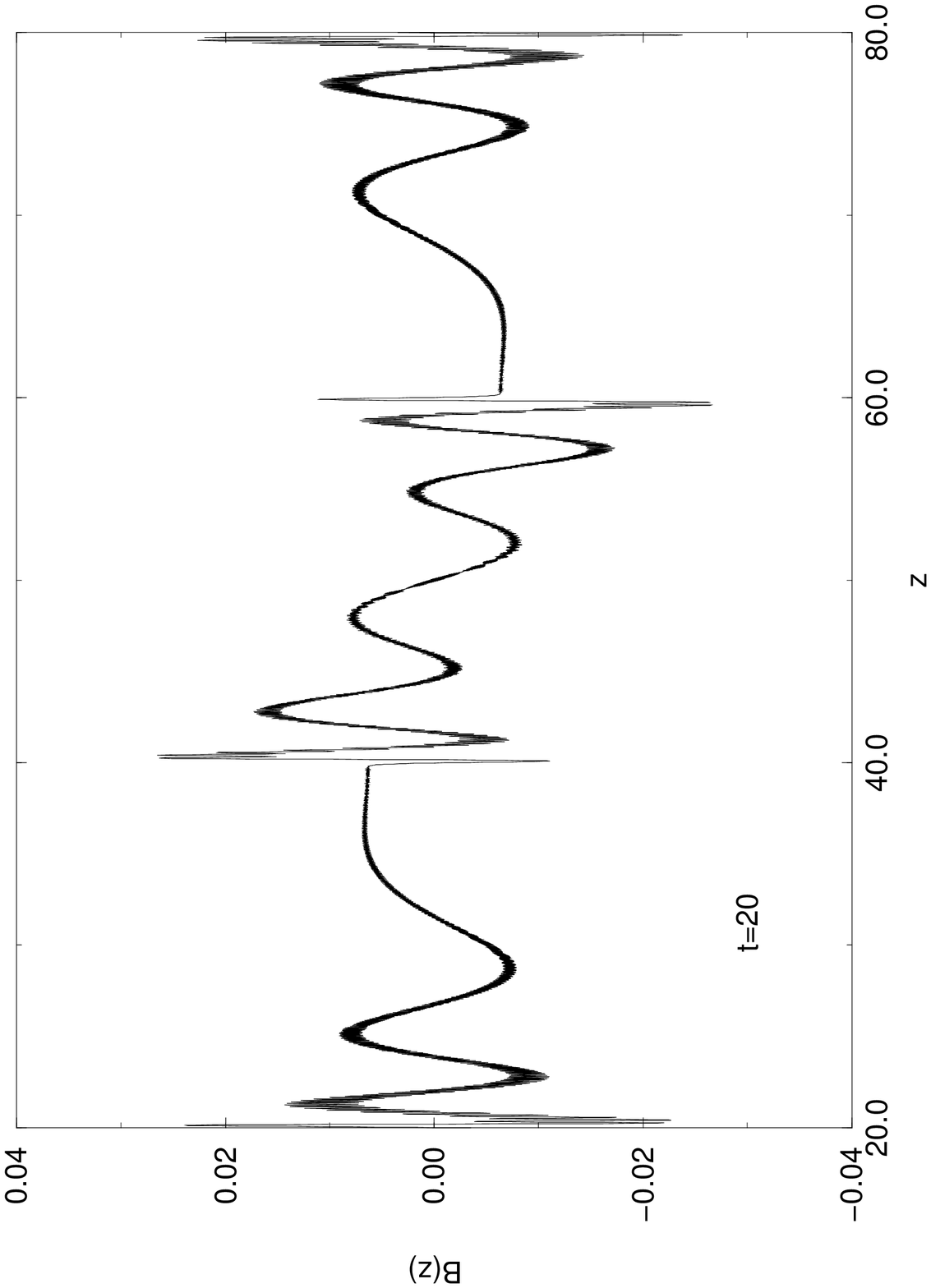}
\includegraphics{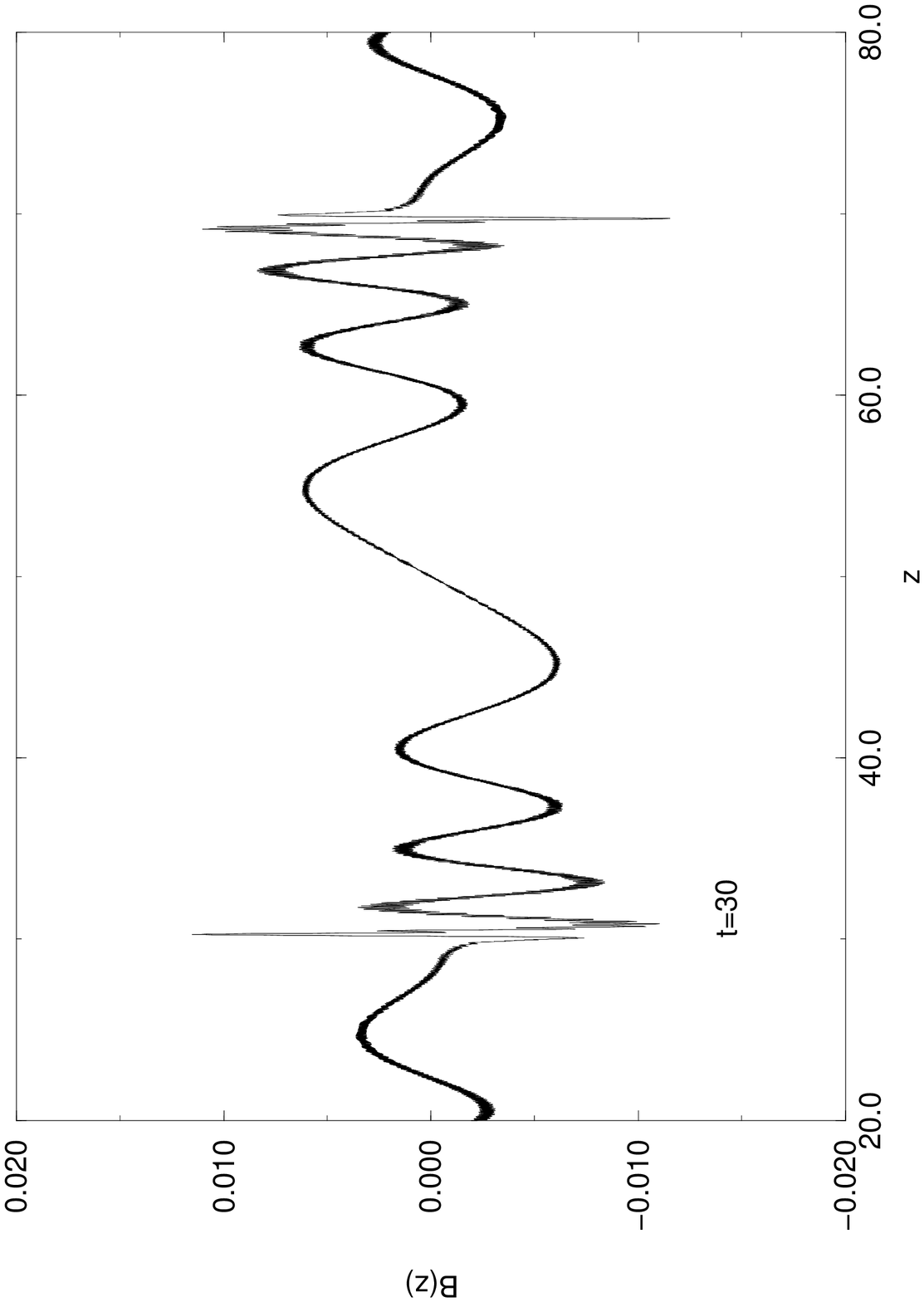}
\includegraphics{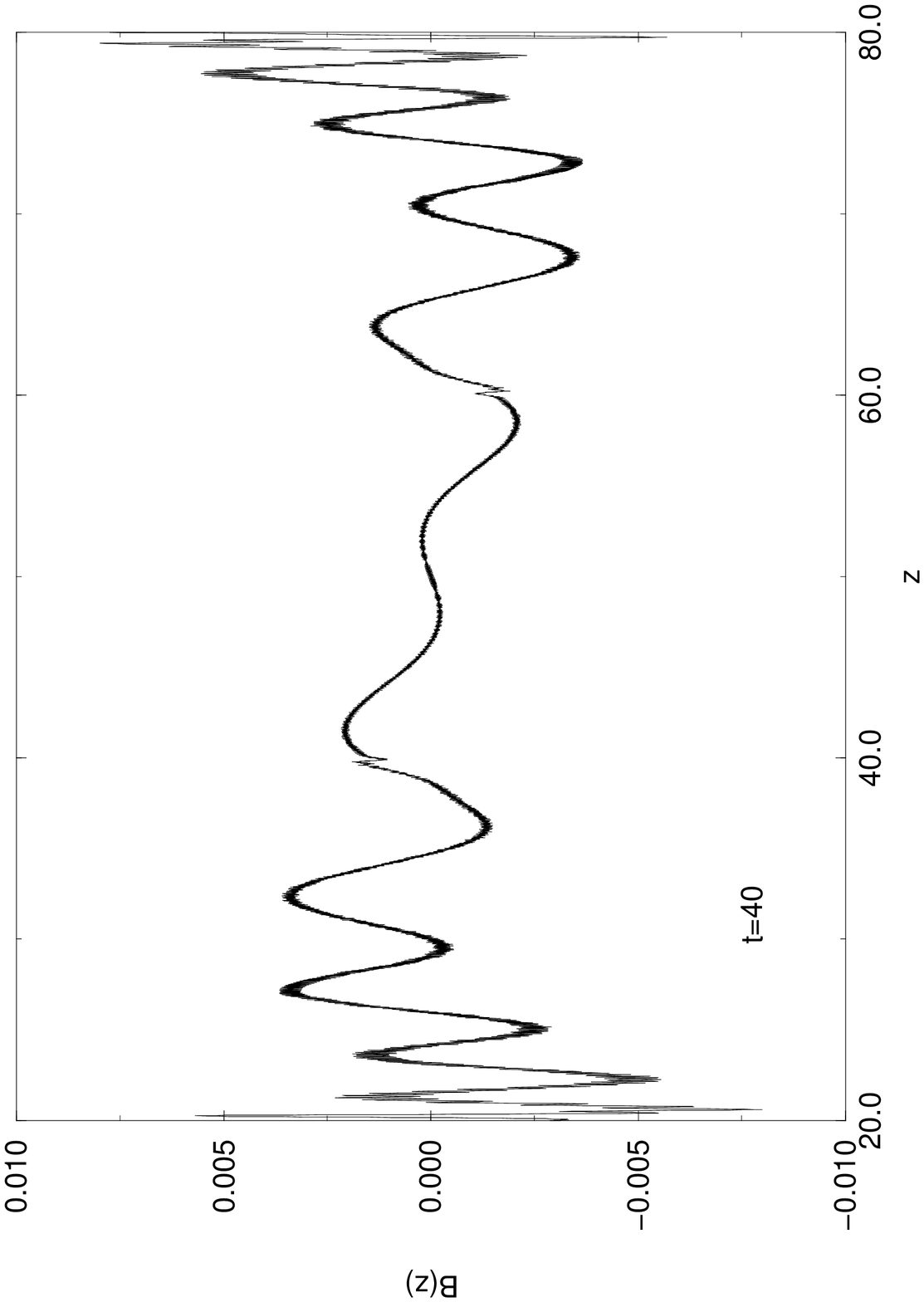}
\caption{The time evolution of $B$ in a symmetric collision 
of three bubbles each having radius
$R=10$ at the collision. Here $r=1$ and time refers to the time elapsed
after the initial collision of the bubbles. The points of initial collision
of the bubbles on the z-axis are 40 and 60 (and units are $e\eta=1$).}
\label{kuva4}
\end{figure}

A simple way to discuss the collision of several bubble is to 
assume that one of the bubbles is much larger than the others, as was
done in Sect. 2.2. 
When $\tau\rightarrow\infty$, the equation of motion is approximately
the wave equation, and thus the small bubbles only create small
wave-like perturbations which subsequently interfere with each other.
Our conclusion is that, qualitatively, all possible collisions between
the bubbles look like collisions of empty bubbles. 

\section{Colliding electroweak bubbles}
\subsection{Flux spreading and diffusion}
The magnetic field  generated in bubble collisions
will be imprinted on the background plasma.
In the early universe electrical conductivity is high but not
infinite. When $1\GeV \ll T \lsim m_W
$ it is given by \cite{us}
\be{conduct}
\sigma\simeq 6.7T
\ee
When $T\gg m_W$, and above the electroweak phase transition, one should also
account for the $W$, $Z$ and the higgs, but their presence will change
$\sigma$ only slightly (note that quarks, including the top, contribute very
little \cite{us} to $\sigma$). 
Therefore we will adopt \eq{conduct} as appropriate
for electrical conductivity both in the broken and in the
unbroken phase. Finite conductivity
gives rise to a diffusion term in the MHD equation
\be{mhd}
{\partial \vec B\over \partial t}=\nabla\times(\vec v\times\vec B)
-\frac 1\sigma \nabla\times(\nabla\times \vec B)
\ee
so that diffusion begins with small scales, and diffusion time 
$t_d\simeq \sigma L^2$. 

\begin{figure}
\leavevmode
\centering
\vspace*{50mm}
\includegraphics{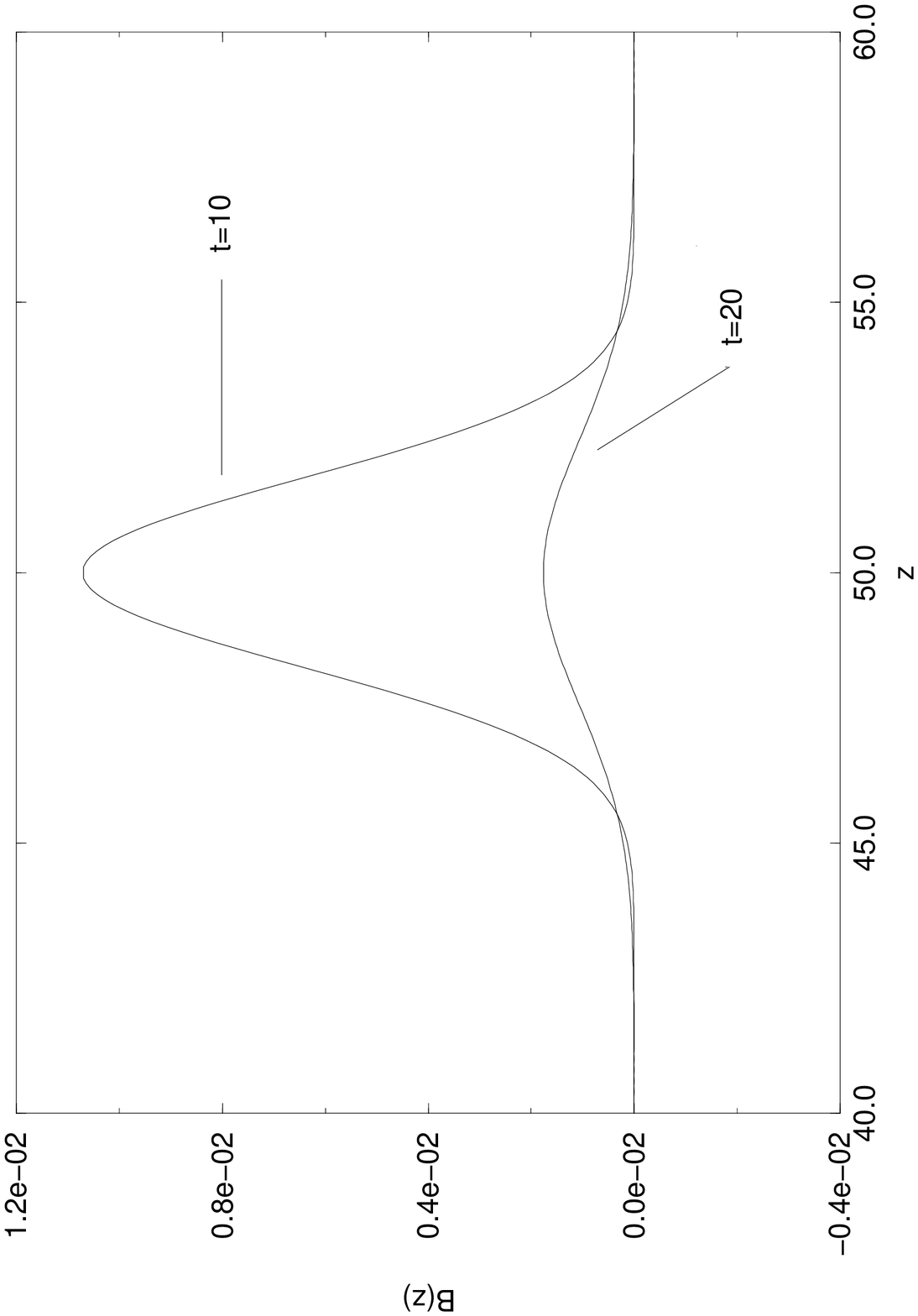}
\includegraphics{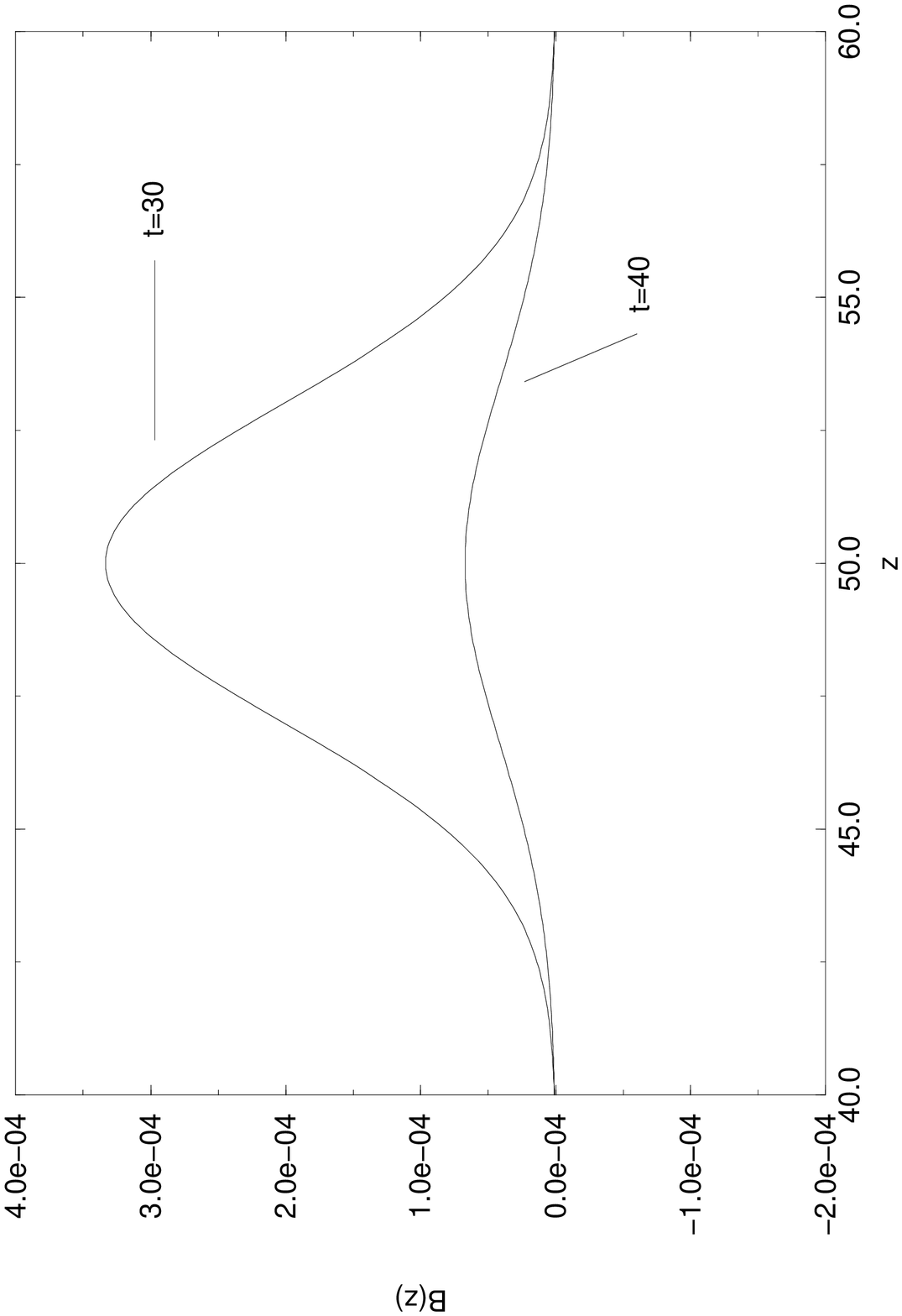}
\caption{$B$ with $v=1$, $r=1$, $\sigma=7$, $R=10$ and $t=10$, $20$, $30$ and
$40$ after the initial collision. The point of initial collision on the z-axis
is $z_1=50$ (and units are $e\eta =1$).}
\label{kuva5}
\end{figure}

In real cosmological phase transitions the velocity of the bubble wall
is definitely less than $c$.  In such a situation
the magnetic flux will escape the region of intersection and penetrate
the interior of the bubbles, and eventually the false vacuum outside
\cite{kv}.
This issue can be addressed by taking the bubble radius to be $R(t)=vt$
and assuming $r\equiv x^2+y^2 \ll 1$ so that we may still take 
$\tau\equiv\sqrt{v^2t^2-r^2}$. (Here we assume that
the nucleated bubbles have the same radius.) 
To include dissipation in the equation of motion properly, we add a
conduction current term $j^{\mu}_{c}$ to the Maxwell's equations 
\cite{kv} and
obtain the following equation of motion for $f(\tau,z)$:
\bea{eqmsmallv}
&-&[\frac{r^2-v^2(\tau^2+r^2)}{\tau^2}]\frac{\partial^2f}{\partial\tau^2}
+(\frac{2}{\tau}-\frac{1}{\tau^3}\left\{ 
r^2(1+v^2)-2 (\tau^2+r^2)\right\}
\nonumber \\
&+& 
\sigma\frac{(1+v^2)\sqrt{\tau^2 +r^2}}{2v\tau})
\frac{\partial f}
{\partial\tau}-
\frac{\partial^2f}{\partial z^2}
+e^2\eta^2f=0.
\eea
At the time of intersection, $\tau =R$, 
with $R\gg 1/T$, so that ${1}/{\sigma \tau}\simeq {1}/{\sigma R}\ll 1$.
Neglecting higher order terms results in
\be{eqm2smallv}
\left(\frac{v^2\partial^2}{\partial \tau^2}+\frac{\sigma (1+v^2)}{2v}\frac{\partial}{
\partial \tau}-\frac{\partial^2}{\partial z^2}+e^2\eta^2\right)f=0~.
\ee
The consequence of the diffusion term in \eq{eqm2smallv}
is to smooth out the rapid oscillations
of $\vec B$, as is demonstrated in Fig. 5, where we display
the numerical solution of \eq{eqm2smallv} for the case $v=1$. 
Fig. 5
should be compared with Fig. 2, which does not include diffusion.

As will be discussed below, in electroweak
phase transition the bubbles will in fact intersect with non-relativistic
velocities. In Fig. 6 we show the velocity effect
on the created magnetic field. As can be seen from Fig. 6
(and comparing with Fig. 5), the 
smaller the velocity, the smaller the magnetic field. 

\begin{figure}
\leavevmode
\centering
\vspace*{50mm}
\includegraphics{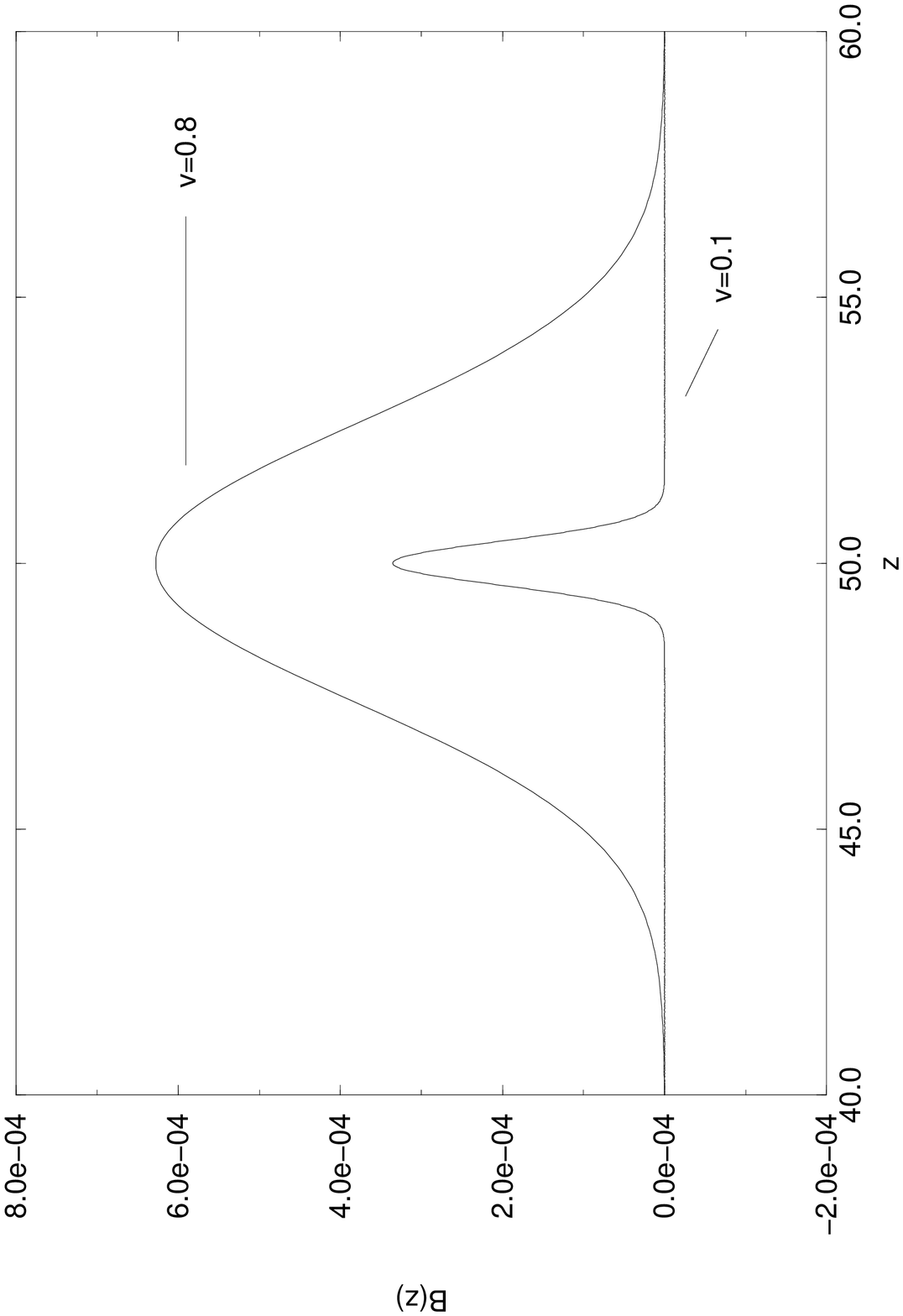}
\includegraphics{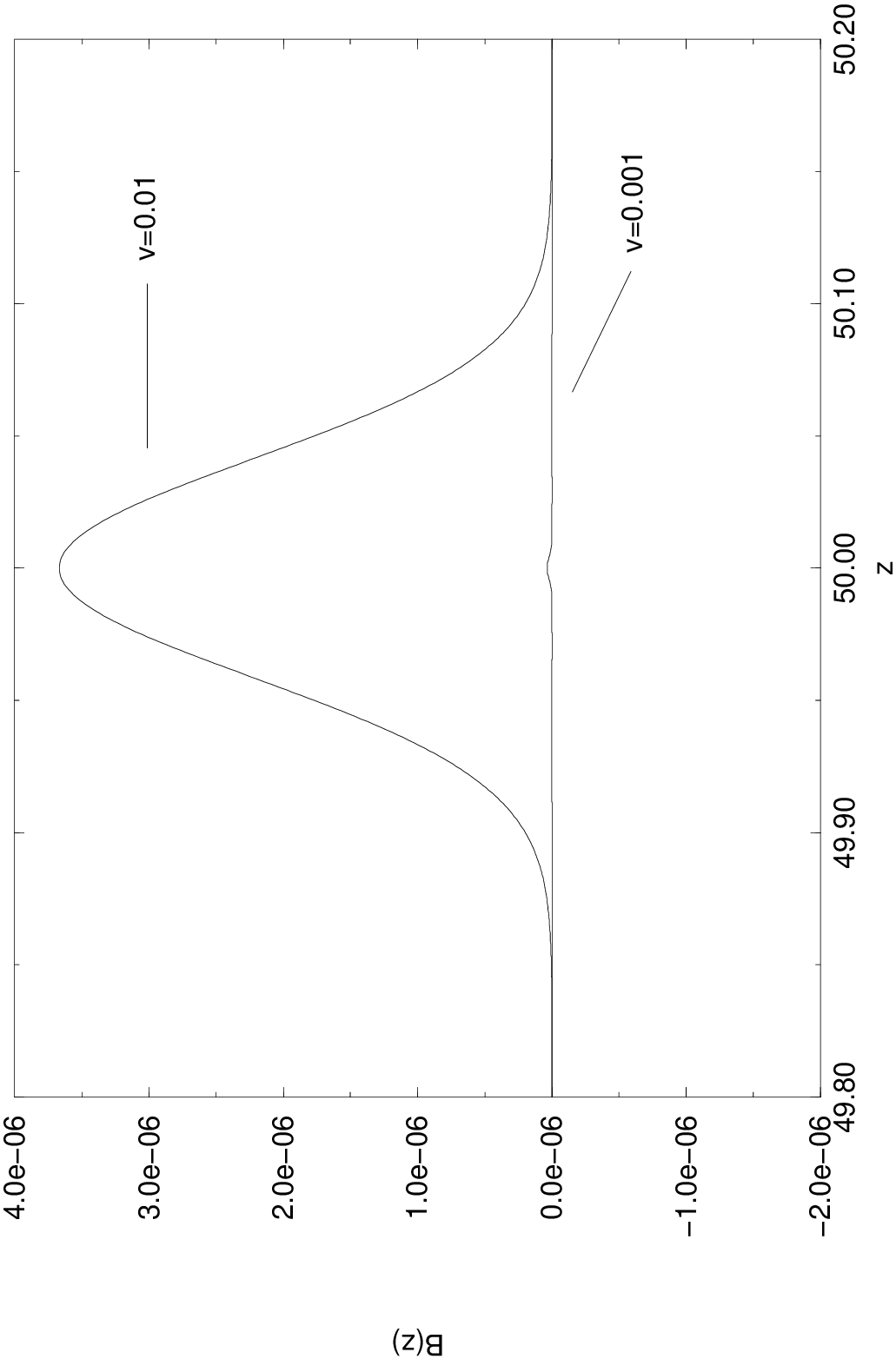}
\caption{$B$ with $r=1$, $R=10$, $\sigma =7$, $t=30$ after the initial 
collision and
$v=0.8$, $0.1$, $0.01$ and $0.001$. The point of initial collision on the 
z-axis is $z_1=50$ (and units are $e\eta =1$).} 
\label{kuva6}
\end{figure}

Fig. 6 does not yet take fully into account the MHD equation, which 
should hold also outside the intersection region. Taking \eq{mhd}
in conjunction with \eq{eqmsmallv} results in
the escape of the magnetic flux, which is most easily seen 
in  the case of low conductivity ($\sigma=0$) and is demonstrated in
Fig. 7. In Fig. 8  magnetic diffusion is again switched on.
Again the magnetic field escapes
from the intersection region and moves outwards with the speed of light.
 Here our conclusion is different from \cite{kv}, where it was argued
that with bubble wall 
velocities in the range
$v\simeq 0.1-1$ no magnetic field should be present outside the
collision region. 
For very low velocities ($v\leq 0.01$)
with $\sigma=7$
the magnetic field looks however just like the case with $\sigma=0$. This is
due to the fact that for small bubble wall velocity the 
magnetic field outside the bubbles
has enough time to decay. Note, however, that decay time is dependent on the
scale of magnetic structure, i.e. the radii of the bubbles, which in
these examples are very small compared to the realistic EW case (
the reason being that bubbles
with very large radii are not easily tractable by numerical analysis and
that in the region of interest, $\tau\simeq R$, the structure 
of the magnetic field does not depend essentially on $R$).

\begin{figure}
\leavevmode
\centering
\vspace*{95mm}
\includegraphics{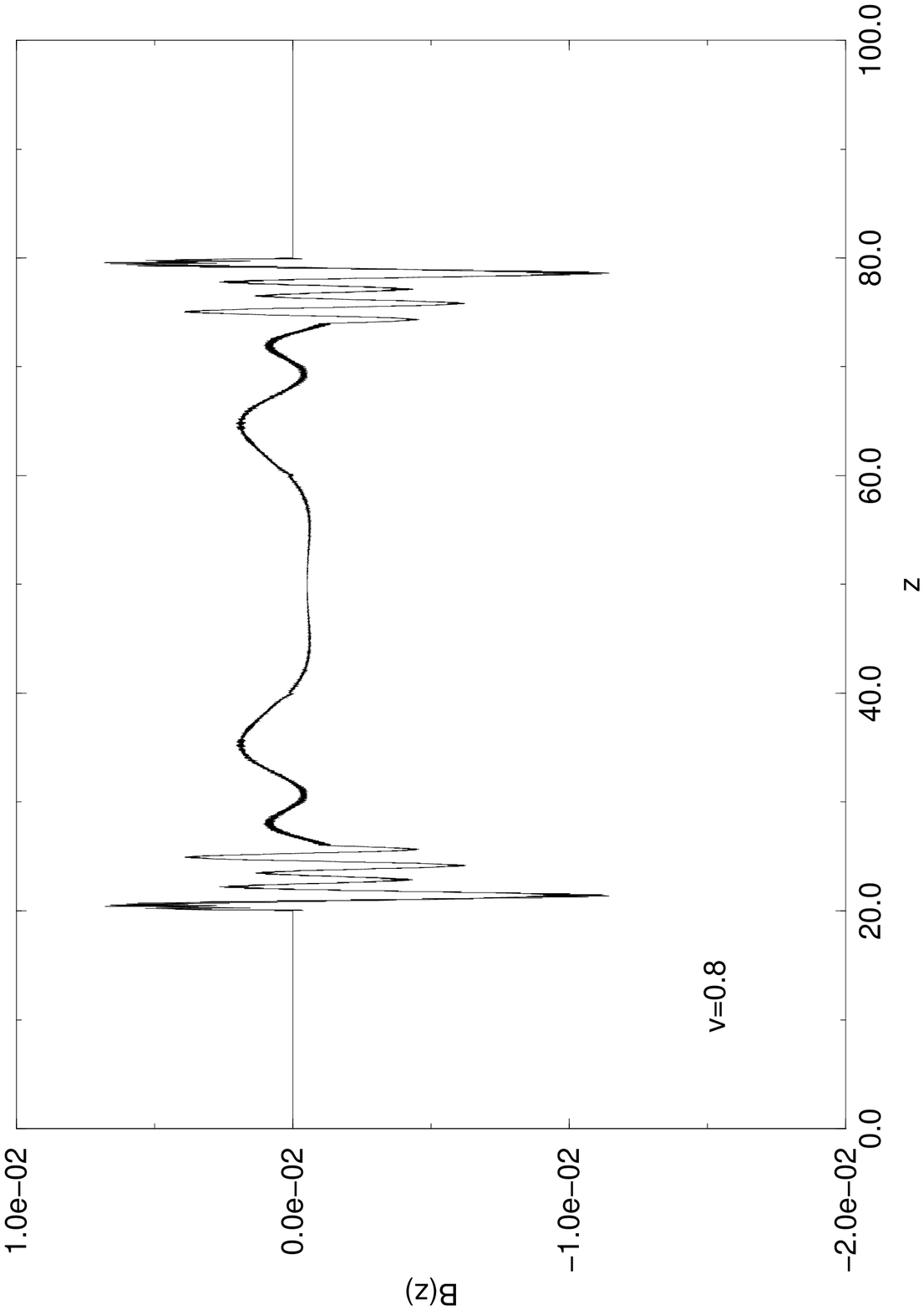}
\includegraphics{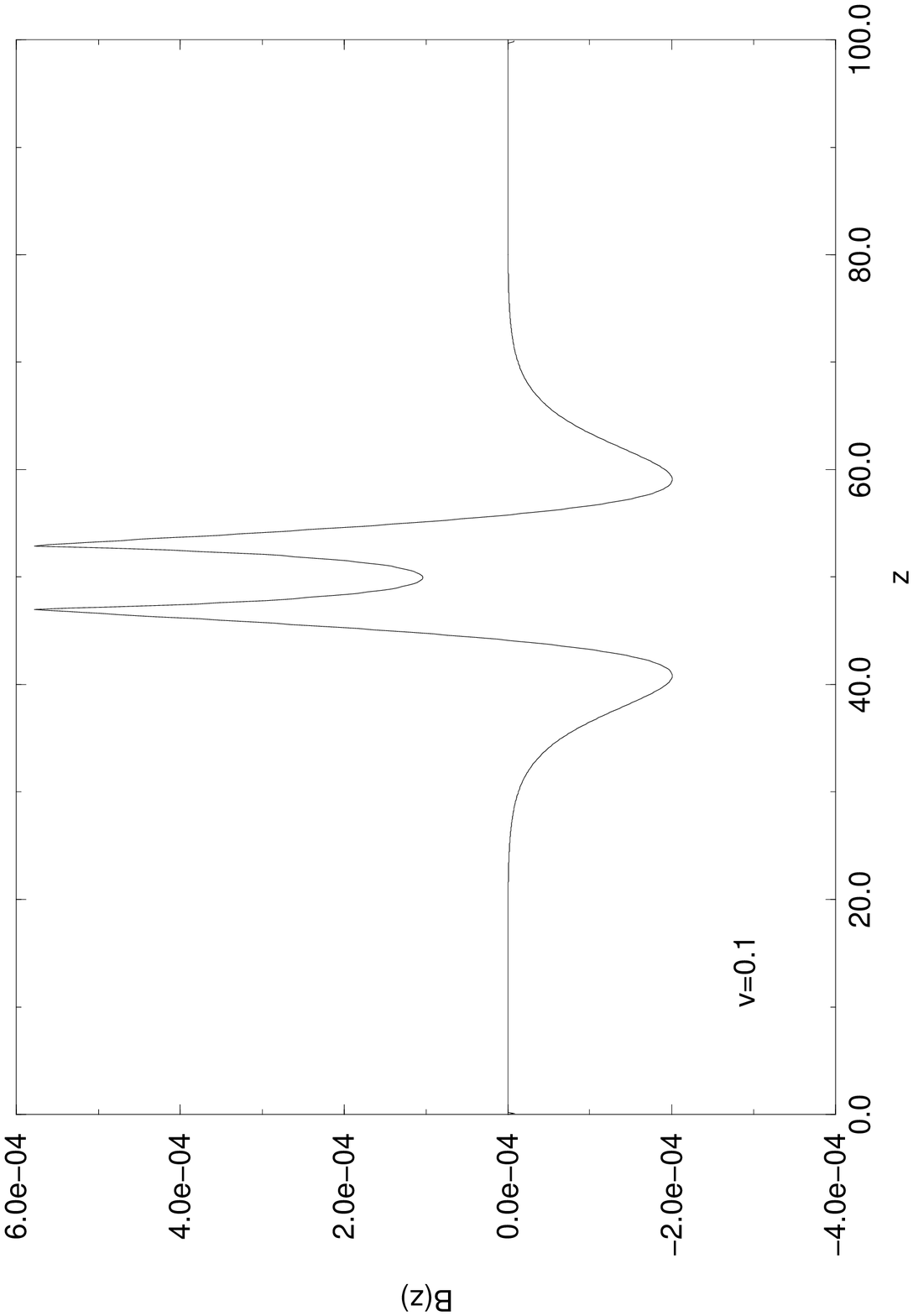}
\includegraphics{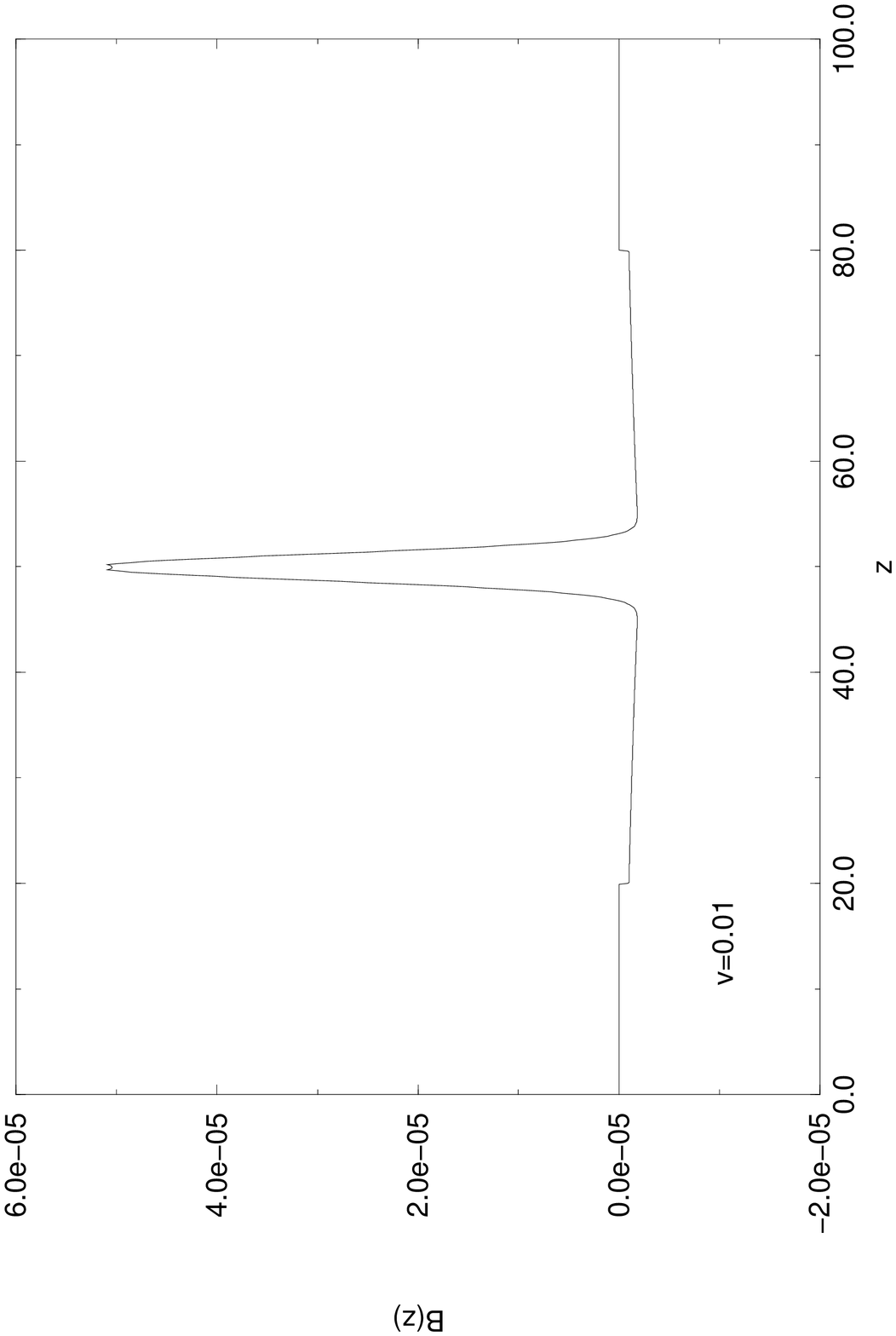}
\includegraphics{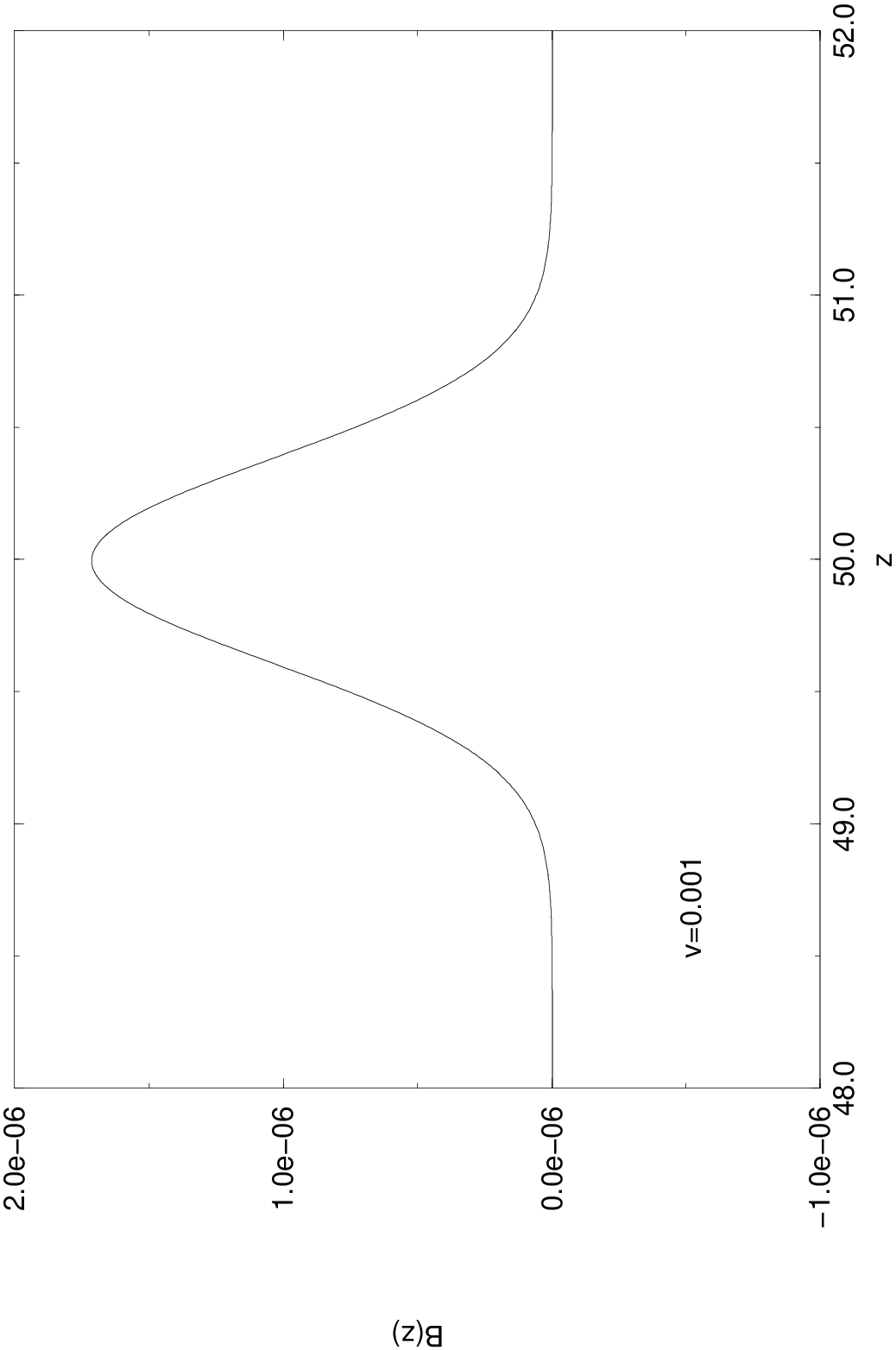}
\caption{$B$ with $r=1$, $R=10$, $\sigma =0$, $t=30$ after collision and
$v=0.8$, $0.1$, $0.01$ and $0.001$. The point of initial collision on the 
z-axis is $z_1=50$, and the outer edge of the intersection region is at 
$\simeq 50\pm vt$. 
The outer edge of the magnetic field is at $\simeq 50\pm t$ 
(and units are $e\eta =1$).} 
\label{kuva7}
\end{figure}

\begin{figure}
\leavevmode
\centering
\vspace*{45mm}
\includegraphics{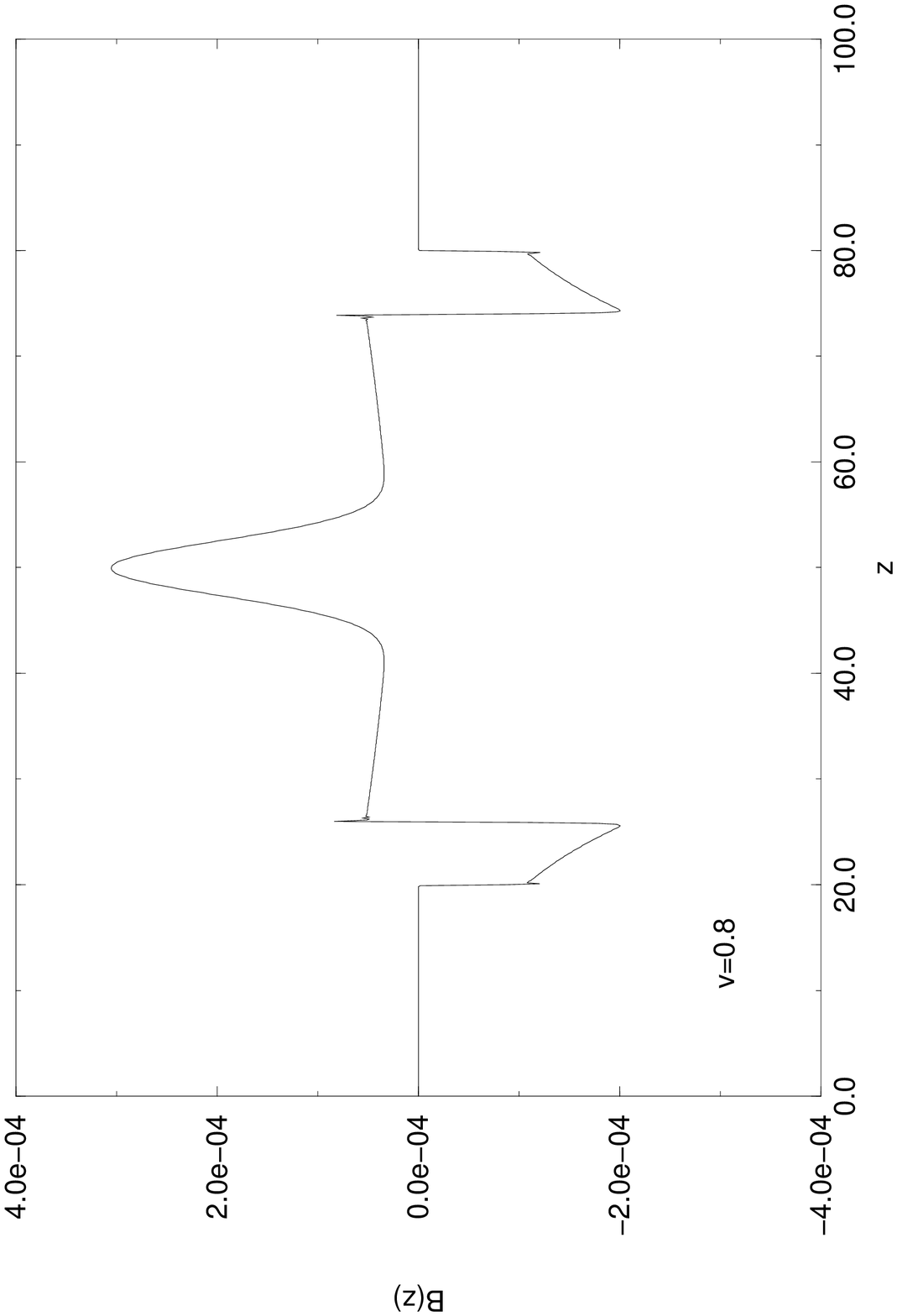}
\includegraphics{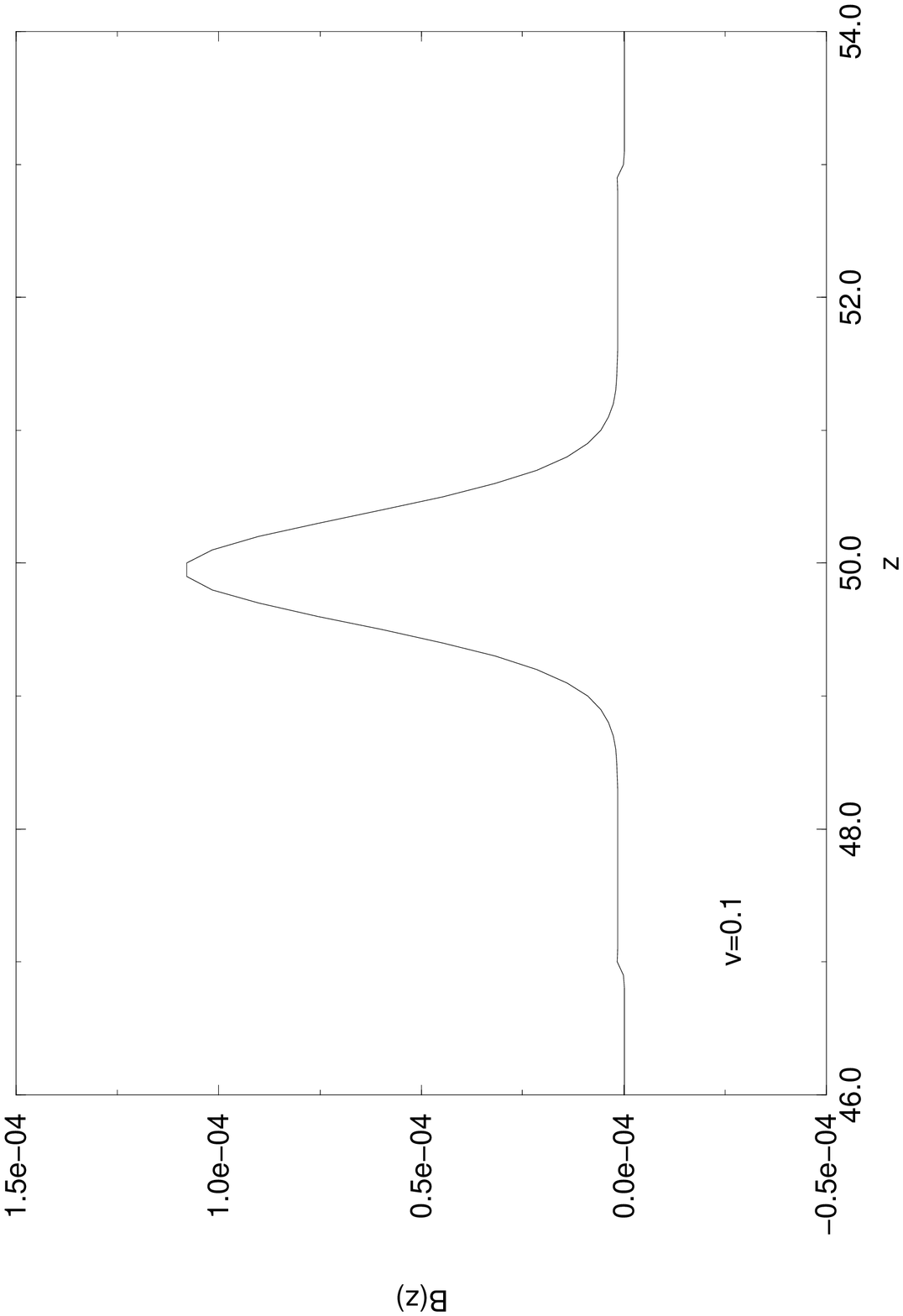}
\caption{$B$ along $r=1$ with $R=10$, $\sigma =7$ and 
$t=30$ after collision with
$v=0.8$ and $0.1$. For the cases  $v=0.01$ and $0.001$ 
the field looks
like in Fig. 6. The point of initial collision on the 
z-axis is $z_1=50$, and the outer edge of the intersection region is at 
$\simeq 50\pm vt$. The outer edge of the magnetic field is at $\simeq 50\pm t$ 
(and units are $e\eta =1$).}  
\label{kuva8}
\end{figure}

\subsection{Electroweak phase transition}
The real-time history  of a first order electroweak phase transition 
depends in an essential way on the hydrodynamical, non-equilibrium
dynamics. Recent lattice simulations \cite{lattice} have now provided
reliably the parameters pertaining the phase transition
in the case $m_H\lsim m_W$, and a
real-time simulation of the bubble growth and collision
has been performed in \cite{hannu}. In the following we shall make use
of these results.

When bubbles of broken symmetry are first nucleated, they
usually grow rapidly. In the electroweak case the initial growth of the
bubble wall is by subsonic deflagration, with velocities of the order 
of 0.05$c$,
depending on the assumed friction strength. The wall is preceded by a shock
front, which may collide with other bubbles. This results in reheating,
oscillations of the bubble radii, but
eventually a phase equilibrium is attained. The ensuing bubble growth is
very slow and takes place because of the expansion of the universe
\cite{hannu}. Note that because the universe has been reheated back to $T_c$,
no new bubbles are formed during the slow growth phase.

As an example, let us choose as the reference point the set of numbers related
to the case of a small friction coefficient
\cite{hannu}:
\be{numerot}
r_{ave}=9.5\times 10^{-8}t_H~;~~v=1.2\times 10^{-4}~;~~\delta=0.03~~,
\ee
where $r_{ave}$ is the average distance of the nucleation centers, $v$ is the
velocity of the bubbles after the collision of the shock fronts,
and $\delta$ is the fraction of the volume in the broken phase at
the onset of the slow bubble expansion (the initial velocity 
of the wall in
this case is $v_{init}=0.089$). The numbers refer to the case of
$m_H=68\GeV$, a weak first order transition. To be definite, we
will adopt $T_c=100\GeV$ and $t_H=3.55\times 10^{13}\GeV^{-1}$.

In EW phase transition the collision of the bubbles thus takes 
place with a very slow velocity, and the bubbles are very large. 
Therefore the discussion in the previous Section is not directly
applicable, but we may nevertheless assume that the gross features
hold, in particular that the collision between EW bubbles will
(for practical purposes) always resemble the collision of two empty
bubbles.

In Fig. 6 we displayed the numerical solution to the equation of
motion \eq{eqm2smallv}. For large EW bubbles numerical methods
are not accurate enough to provide a reliable result. Therefore
we need an analytic esimate of \eq{eqm2smallv}. Let us write
$f(\tau,z)=b(\tau)c(z)$ and use the initial conditions
$f(R,z)=0$ and $\partial_{\tau} f(R,z)={e\eta^2\Theta_{0}\cdot
\epsilon (z)}/{R}$, where $\epsilon (z)$ is the sign-function. Then in 
the small $r\equiv\sqrt{x^2+y^2}$ limit we find that 
\bea{bsmallv}
B&=&4\frac{e\eta^2\Theta_0v^3r}{\pi R}\mbox{e}^{\frac12{\sigma (1+v^2)(R-\tau )
}/{2v^3}} \nonumber \\ 
&\times&\int_{-\infty}^{\infty}\cos (k(z-z_1))\frac{\sinh (\frac12{\sqrt{\sigma^2 
(1+v^2)^2-
16q^2v^4}(\tau -R)}/{2v^3})}{\sqrt{\sigma^2 (1+v^2)^2-16q^2v^4}}dk~,
\eea
where $z_1$ is the point of initial collision on the z-axis, $q^2\equiv 
(k^2+e^2\eta^2)$ and $\tau^2=v^2t^2-r^2$.

From \eq{bsmallv} one sees that $B$ at extremely small scales  
($L\lsim 4v/\sigma=7.1\times 10^{-5}/T$) decays
essentially instantaneously, although strictly speaking one cannot
apply MHD at length scales less than the interparticle distance $\sim 1/T$.
At scales $L\gsim 1/T $ \eq{bsmallv} implies that at a given time $t_d\gsim
R/v$ the magnetic field at length scales 
\be{scales}
L\lsim \left({\sqrt{R^2+r^2}-R\over v\sigma}\right)^{1/2}
\ee
has already decayed. Note that at the outer edges of the intersection
region diffusion cannot remove $B$, unless $r\simeq 0$, $L\ll R$.

Roughly speaking, at the onset of the slow bubble expansion, the bubbles
almost touch. Once they have collided, the phase transition will be
completed in a time $\Delta t\simeq {\cal O}(R/v)$.
Although strictly speaking \eq{bsmallv} is valid only for $r\ll vt$, 
let us use it 
to estimate the magnitude of $B$ at the end of phase transition.
From \eq{scales} we can see that the magnetic field has decayed from the
centre of the collision region. \eq{bsmallv} tells us that new magnetic field 
is created only in the region where $\tau\simeq R$. The flux
escapes, as discussed in Sect. 3.1, and dissipation
outside the bubbles is very slow (at scales much larger than
$L$ in \eq{scales}). Inside the bubbles there is also a region where 
$B$ has not yet decayed. The largest field is obtained, however,
around $\tau\simeq R$, so that 
 the field looks like a 
narrow ring around the z-axis at $z=0$ and $r=\sqrt{v^2t^2-R^2}$.
From \eq{bsmallv} we then obtain an estimate for the strength of the magnetic 
field in the ring
\bea{approsmallv}
B(R)\simeq\frac{e\eta^2 v}{R}\sqrt{\gamma^2+2\gamma R},
\eea
where we have defined $\gamma\equiv v\Delta t={\cal O}(1) R$. 
This yields  the size and the coherence length
$L_0$ of the field as
\bea{coher}
B & \simeq & 4.0\sqrt{\gamma^2+2\gamma R}/R\GeV^2=2.0\times 10^{20}
\sqrt{\gamma^2+2\gamma R}/R \, 
\G, \nonumber \\
L_0 & \simeq & 6.5\times 10^5((1+\frac{\gamma^2+2\gamma R}{R^2})^{1/2}-1)^{1/2}
/T_c~,
\eea
where we have assumed $\Theta_0=1$, $T_c=e\eta=100\GeV$ and used \eq{numerot}.

Thus we may argue that at 
the end of EW phase transition there are about  $(t_H/r_{ave})^3\sim
10^{21}$ magnetic rings within each horizon volume. To obtain a
reliable estimate of the average $B$, we should average over the
possible bubble sizes, on which $B$ depends.
The spectrum of separation of the adjacent shocked spherical bubbles
in a first order phase transition has been estimated in \cite{meyer} and reads
\be{distrib}
P(R)=\frac{96}{185}\left({S'\over v_s}\right)^3R^2
\Bigl[\exp(-S'R/v_s)\bigl({S'R\over 2v_s}-\frac23\bigr)
+\exp(-2S'R/v_s)\bigl({S'R\over 4v_s}+\frac23\bigr)\Bigr]~,
\ee
where 
$v_s=1/\sqrt{3}$ is the velocity of the shock front, and
\be{spilkku}
S'/v_s\equiv S'(t_f)/v_s=(\pi)^{1/3}/r_{ave}=4.35\times 10^{-7}\GeV
\ee
is the derivative of the tunneling action at nucleation time $t_f$
\cite{janne},
and the number is for the reference values \eq{numerot}. Assuming that
\eq{distrib} also gives the distribution of the bubble radii at
collision, we thus arrive at the average magnetic field
\be{average}
\langle B\rangle = \int P(R)B(R)dR=3.1\times 10^{20} \G ,
\ee 
where we have taken $\gamma=2R$ for definitess.

Thus, on the average, each volume of a radius $r_{ave}$ contains a
large ring-like field $\langle B\rangle$, but with the planes of inclination
randomly distributed. For large volumes, we should average over all
possible inclinations, which corresponds to a random walk on the
2d surface of a sphere. This results in a highly entangled field
with a root-mean-square value
$
\langle {\bf B}_{rms}\rangle = 
\sqrt{\sum_{n,k=1}^{N}\langle {\bf B}_n\rangle\langle {\bf B}_k\rangle/N^2} =
\langle B\rangle L_0/L.
$
Therefore, the rms-field at a given time and comoving scale $L$ reads
\be{rms}
B_{rms}(t,L)=f_T(t)\langle B\rangle \left({R_{EW}\over R(t)}\right)^2
\frac{L_0(t)}{L}=
f_T(t)\langle B\rangle \left({R_{EW}\over R(t)}\right)
\left({r_{ave} \over L}\right)~,
\ee
where the factor $f_T(t)$ accounts for the turbulent enhancement of
${\bf B}$ at large length scales, which persists until the
plasma becomes matter dominated and even in the
presence of large plasma viscosity (i.e. Silk damping) \cite{axel}.
The reason for such an inverse energy cascade is the non-linear nature
of the MHD equations. Effectively, the small magnetic loops will merge
to form larger magnetic loops, thereby transferring energy to larger
length scales.
Numerical simulations, using so-called shell models to simulate the
full 3d MHD equations, together with 
scaling arguments, suggest that $f_T(t)\simeq t^p$ with $p\simeq 0.25$.
The total enhancement may thus be estimated as $f_T\simeq 10^6$ (but
with somewhat large uncertainty).
At the comoving scale
of 10 Mpc today we thus find for the set of the reference values
\eq{numerot} (which implicitly assume that $m_H=68$ GeV)
\be{tulos}
B_{rms}\simeq 10^{-21} {\rm G}~.
\ee
(Here we assumed that the transition from
radiation dominated to matter dominated era takes place at 
$t\simeq 10^5$ yrs). It is encouraging that the magnetic field 
given by \eq{tulos} appears to be of the correct order of
magnitude to provide  the seed field for the galactic dynamo.
\section{Discussion}
Although we have completely neglected the non-abelian nature of
the electroweak theory, it is unlikely to affect the gross features
of true electroweak bubble collisions. They, like in the abelian
Higgs model we have considered, are bound to depend mostly on
the existence of a phase difference and magnetic diffusion
(although the abelian anomaly could be important for
magnetic fields \cite{shapo}). 
We have not treated the full (and complicated)
hydrodynamics of the bubbles either. These
may include deformations of the spherical bubbles and kinematic
viscosity playing an important part in the dynamical evolution. 
If the surface tension of
the electroweak bubbles is not extremely low, sphericality should
be a good approximation even during  bubble collisions. 
The resulting magnetic fields, although not necessarily perfectly 
ring-like, should be of the same order of magnitude as discussed
here. For very low surface tension the situation might be
different.

Our treatment is obviously a simplification of the highly complex
chain of bubble collisions. However, it serves to emphasize the  
decisive role  the velocity of the colliding bubbles play in
magnetic field generation. Velocity itself depends on 
hydrodynamical details such as the strength of the shock front,
and in the final analysis on the parameters of the Higgs potential,
in particular on the mass of the higgs. Increasing the higgs mass
decreases the velocity of the colliding bubbles so that
the magnetic field would also decrease according to
\eq{approsmallv}. For example, moving
from $m_H=50$ GeV to $m_H=68$ GeV entails a reduction of $v$ by a factor of
ten \cite{hannu}. At the same time the surface tension decreases by a factor of
twenty. It is also known \cite{keijo} that for high enough higgs mass, 
the transition is no longer of first order. It therefore seems
likely that for large enough higgs mass the present treatment
is no longer valid, and no magnetic field is created.

The fact that reasonable
assumptions seem to produce primordial magnetic fields which
could serve as the seed field for the galactic dynamo
nevertheless
emphasizes the significance of the electroweak phase transition.
There now exist proposals to measure the strength of the 
intergalactic magnetic field to a very high precision \cite{plaga}.
It is interesting that such measurements could provide information
also on the nature of the electroweak phase transition.
\section*{Acknowledgments}
We thank Hannu Kurki-Suonio for illuminating discussions, and the Aspen
Center of Physics, where part of the work was done, for hospitality.
This work is partly supported by NorFa and the
Academy of Finland.
\newpage

\end{document}